\documentclass[fleqn,usenatbib,useAMS]{mnras}

\usepackage{newtxtext,newtxmath}
\usepackage[T1]{fontenc}
\usepackage{amsmath}
\usepackage{xcolor}
\usepackage{xspace}
\usepackage{xurl}
\usepackage{array}
\usepackage{graphicx}
\usepackage{orcidlink}
\usepackage{longtable}
\usepackage{booktabs}

\newcommand{\kms}{\,km\,s$^{-1}$\xspace}
\newcommand{\alphavir}{\alpha_\mathrm{vir}}

\title[ACES overview]{ALMA Central Molecular Zone Exploration Survey (ACES) I: Overview}

\newcounter{affcounter}

\newcommand{\defaffiliationlabel}[1]{%
  \refstepcounter{affcounter}%
  \expandafter\xdef\csname #1\endcsname{\theaffcounter}%
}

\newcommand{\affref}[1]{$^{\csname #1\endcsname}$}

\newcommand{\affrefs}[1]{%
  $^{%
    \@for\@ref:=#1\do{%
      \@ref\@ifnextchar\@nil{}{,}%
    }%
  }$%
}

\newcommand{\affrefTwo}[2]{$^{\csname #1\endcsname,\csname #2\endcsname}$}
\newcommand{\affrefThree}[3]{$^{\csname #1\endcsname,\csname #2\endcsname,\csname #3\endcsname}$}
\newcommand{\affrefFour}[4]{$^{\csname #1\endcsname,\csname #2\endcsname,\csname #3\endcsname,\csname #4\endcsname}$}

\newcommand{\printaffiliation}[2]{%
  $^{\csname #1\endcsname}$#2\\%
}
\defaffiliationlabel{ljmu}
\defaffiliationlabel{COOL}
\defaffiliationlabel{colorado}
\defaffiliationlabel{eso}
\defaffiliationlabel{uconn}
\defaffiliationlabel{cab_csic}
\defaffiliationlabel{uflorida}
\defaffiliationlabel{mpia}
\defaffiliationlabel{iaa_taipei}
\defaffiliationlabel{kansas}
\defaffiliationlabel{chalmers}
\defaffiliationlabel{clap}
\defaffiliationlabel{iop_epfl}
\defaffiliationlabel{ukarcnode}
\defaffiliationlabel{mpifr_bonn}
\defaffiliationlabel{kau_jeddah}
\defaffiliationlabel{princeton}
\defaffiliationlabel{armagh}
\defaffiliationlabel{nrao}
\defaffiliationlabel{mpe}
\defaffiliationlabel{ita_heidelberg}
\defaffiliationlabel{unknown_affil}
\defaffiliationlabel{uva}
\defaffiliationlabel{upenn}
\defaffiliationlabel{leiden}
\defaffiliationlabel{anu}
\defaffiliationlabel{iaa_csic}
\defaffiliationlabel{malta}
\defaffiliationlabel{jbca}
\defaffiliationlabel{cassaca}
\defaffiliationlabel{ucn}
\defaffiliationlabel{csic_gen}
\defaffiliationlabel{heidelberg_univ}
\defaffiliationlabel{northwestern}
\defaffiliationlabel{shao}
\defaffiliationlabel{naoj}
\defaffiliationlabel{ias}
\defaffiliationlabel{ucl}
\defaffiliationlabel{mit}
\defaffiliationlabel{izw_heidelberg}
\defaffiliationlabel{cfa}
\defaffiliationlabel{radcliffe}
\defaffiliationlabel{uw_madison}
\defaffiliationlabel{uw_seattle}
\defaffiliationlabel{ari_heidelberg}
\defaffiliationlabel{naoc_key}
\defaffiliationlabel{ucas_astro}
\defaffiliationlabel{eso_chile}
\defaffiliationlabel{jao}
\defaffiliationlabel{ipm}
\defaffiliationlabel{ucla}
\defaffiliationlabel{standrews_eso}
\defaffiliationlabel{keio_univ}
\defaffiliationlabel{inaf_gen}
\defaffiliationlabel{nanjing}
\defaffiliationlabel{nanjing_key}
\defaffiliationlabel{umd}
\defaffiliationlabel{ulaserena}
\defaffiliationlabel{ice_csic}
\defaffiliationlabel{ieec}
\defaffiliationlabel{gbo}
\defaffiliationlabel{nice}
\defaffiliationlabel{utokyo}
\defaffiliationlabel{bologna}
\defaffiliationlabel{insubria}
\defaffiliationlabel{tra_bonn}
\defaffiliationlabel{umass}
\defaffiliationlabel{kiaa_pku}
\author[S.~N.~Longmore et al.]{Steven N. Longmore,\affrefTwo{ljmu}{COOL}\orcidlink{0000-0001-6353-0170}
John Bally,\affref{colorado}\orcidlink{0000-0001-8135-6612}
Ashley~T.~Barnes,\affref{eso}\orcidlink{0000-0003-0410-4504}
Cara Battersby,\affref{uconn}\orcidlink{0000-0002-6073-9320}
Laura Colzi,\affref{cab_csic}\orcidlink{0000-0001-8064-6394}
\newauthor
Adam Ginsburg,\affref{uflorida}\orcidlink{0000-0001-6431-9633}
Jonathan D. Henshaw,\affrefTwo{ljmu}{mpia}\orcidlink{0000-0001-9656-7682}
Paul T. P. Ho,\affref{iaa_taipei}\orcidlink{0000-0000-0000-0000}
Izaskun Jim\'enez-Serra,\affref{cab_csic}\orcidlink{0000-0003-4493-8714}
\newauthor
J.~M.~Diederik Kruijssen,\affref{COOL}\orcidlink{0000-0002-8804-0212}
Elisabeth A.C. Mills,\affref{kansas}\orcidlink{0000-0001-8782-1992}
Maya A. Petkova,\affref{chalmers}\orcidlink{0000-0002-6362-8159}
Mattia C. Sormani,\affref{clap}\orcidlink{0000-0001-6113-6241}
\newauthor
Robin G. Tress,\affref{iop_epfl}\orcidlink{0000-0002-9483-7164}
Daniel~L.~Walker,\affref{ukarcnode}\orcidlink{0000-0001-7330-8856}
Jennifer Wallace,\affref{uconn}\orcidlink{0009-0002-7459-4174}
Emad Alkhuja,\affrefTwo{mpifr_bonn}{kau_jeddah}\orcidlink{0000-0001-6199-9848}
Lucia Armillotta,\affref{princeton}\orcidlink{0000-0000-0000-0000}
\newauthor
Nazar Budaiev,\affref{uflorida}\orcidlink{0000-0002-0533-8575}
Rojita Buddhacharya,\affref{ljmu}\orcidlink{0009-0004-0685-7678}
Alyssa Bulatek,\affref{uflorida}\orcidlink{0000-0002-4407-885X}
Michael Burton,\affref{armagh}
Natalie O. Butterfield,\affref{nrao}\orcidlink{0000-0002-4013-6469}
\newauthor
Laura A. Busch,\affref{mpe}
Paola Caselli,\affref{mpe}\orcidlink{0000-0003-1481-7911}
M\'elanie Chevance,\affrefTwo{ita_heidelberg}{COOL}\orcidlink{0000-0002-5635-5180}
Claire Cook,\affref{kansas}
Samuel Crowe,\affref{uva}\orcidlink{0009-0005-0394-3754}
\newauthor
Ana Karla D\'iaz-Rodr\'iguez,\affref{ukarcnode}\orcidlink{0000-0001-9112-6474}
Enrico DiTeodoro,\affref{unknown_affil}\orcidlink{0000-0000-0000-0000}
Simon R. Dicker,\affref{upenn}\orcidlink{0000-0002-1940-4289}
Katarzyna M. Dutkowska,\affref{leiden}\orcidlink{0000-0003-0980-6871}
\newauthor
Adam Fairley,\affref{ljmu}\orcidlink{0000-0000-0000-0000}
Christoph Federrath,\affref{anu}\orcidlink{0000-0002-0706-2306}
Rub\'en Fedriani,\affref{iaa_csic}\orcidlink{0000-0000-0000-0000}
Zi-Xuan Feng,\affref{clap}\orcidlink{0009-0004-0121-1560}
Karl Fiteni,\affrefTwo{clap}{malta}\orcidlink{0000-0002-9409-8322}
Gary Fuller,\affref{jbca}
\newauthor
Pablo Garc\'ia,\affrefTwo{cassaca}{ucn}\orcidlink{0000-0002-8586-6721}
Javier Goicoechea,\affref{csic_gen}
Philipp Girichidis,\affref{ita_heidelberg}
Simon C.~O.~Glover,\affref{ita_heidelberg}\orcidlink{0000-0001-6708-1317}
Mark Gorski,\affref{northwestern}
\newauthor
Savannah R. Gramze,\affref{uflorida}\orcidlink{0000-0002-1313-429X}
Qi-Lao Gu,\affref{shao}\orcidlink{0000-0002-2826-1902}
H. Perry Hatchfield,\affref{unknown_affil}
Christian Henkel,\affref{mpifr_bonn}
Rebecca J. Houghton,\affref{ljmu}
Pei-Ying Hsieh,\affref{naoj}\orcidlink{0000-0001-9155-3978}
\newauthor
Yue Hu,\affref{ias}\orcidlink{0000-0002-8455-0805}
Katharina Immer,\affref{eso}\orcidlink{0000-0003-4140-5138}
Desmond Jeff,\affrefTwo{uflorida}{nrao}\orcidlink{0000-0003-0416-4830}
Janik Karoly,\affref{ucl}
Jens Kauffmann,\affref{mit}\orcidlink{0000-0002-5094-6393}
\newauthor
Ralf S. Klessen,\affrefFour{ita_heidelberg}{izw_heidelberg}{cfa}{radcliffe}\orcidlink{0000-0002-0560-3172}
Mark R. Krumholz,\affref{anu}\orcidlink{0000-0003-3893-854X}
Alex Lazarian,\affref{uw_madison}
Emily M. Levesque,\affref{uw_seattle}\orcidlink{0000-0003-2184-1581}
\newauthor
Fu-Heng Liang,\affrefTwo{ari_heidelberg}{eso}\orcidlink{0000-0003-2496-1247}
Dani Lipman,\affref{uconn}
Xunchuan Liu,\affref{shao}
Xing Lu,\affrefTwo{shao}{naoc_key}\orcidlink{0000-0003-2619-9305}
Qiu-yi Luo,\affrefTwo{shao}{ucas_astro}\orcidlink{0000-0003-4506-3171}
Alessandro Lupi,\affref{unknown_affil}
\newauthor
Laura McCafferty,\affref{ljmu}
S. Mart\'{i}n,\affrefTwo{eso_chile}{jao}\orcidlink{0000-0001-9281-2919}
Farideh Mazoochi,\affref{ipm}\orcidlink{0000-0003-3390-4893}
Mark R. Morris,\affref{ucla}\orcidlink{0000-0002-6753-2066}
Marie Nonhebel,\affref{standrews_eso}
\newauthor
Francisco Nogueras-Lara,\affrefTwo{iaa_csic}{eso}\orcidlink{0000-0002-6379-7593}
Tomoharu Oka,\affref{keio_univ}\orcidlink{0000-0002-5566-0634}
Juergen Ott,\affref{nrao}
Marco Padovani,\affref{inaf_gen}
Xing Pan,\affrefThree{nanjing}{nanjing_key}{cfa}\orcidlink{0000-0003-1337-9059}
\newauthor
Jaime E. Pineda,\affref{mpe}\orcidlink{0000-0002-3972-1978}
Thushara G.S. Pillai,\affref{mit}\orcidlink{0000-0003-2133-4862}
Marc W. Pound,\affref{umd}
Miguel Requena Torres,\affref{unknown_affil}
\newauthor
Denise Riquelme-V\'asquez,\affref{ulaserena}\orcidlink{0000-0001-5389-0535}
V\'ictor M. Rivilla,\affref{cab_csic}\orcidlink{0000-0002-2887-5859}
Galaxy Salo,\affref{ljmu}
\'Alvaro S\'anchez-Monge,\affrefTwo{ice_csic}{ieec}\orcidlink{0000-0002-3078-9482}
\newauthor
Miriam G. Santa-Maria,\affref{uflorida}\orcidlink{0000-0002-3941-0360}
Rainer Schoedel,\affref{iaa_csic}
Anika Schmiedeke,\affref{gbo}\orcidlink{0000-0002-1730-8832}
Matthias Schultheis,\affref{nice}
\newauthor
Howard A. Smith,\affref{cfa}
Yoshiaki Sofue,\affref{utokyo}\orcidlink{0000-0002-4268-6499}
Leonardo Testi,\affref{bologna}
Grant R. Tremblay,\affref{cfa}\orcidlink{0000-0002-5445-5401}
Arianna Vasini,\affref{insubria}
\newauthor
Gijs Vermari{\"e}n,\affref{leiden}\orcidlink{0000-0002-4346-5858}
Alexey Vikhlinin,\affref{cfa}\orcidlink{0000-0001-8121-0234}
Serena Viti,\affrefThree{leiden}{tra_bonn}{ucl}\orcidlink{0000-0001-8504-8844}
Q. Daniel Wang,\affref{umass}\orcidlink{0000-0002-9279-4041}
Fengwei Xu,\affrefTwo{mpia}{kiaa_pku}\orcidlink{0000-0001-5950-1932}
\newauthor
Suinan Zhang,\affref{nanjing}
Qizhou Zhang,\affref{cfa}\orcidlink{0000-0003-2384-6589}
\\
$^{*}$Author affiliations are listed at the end of the paper
}

\date{Accepted XXX. Received YYY; in original form ZZZ}

\pubyear{2025}

\begin{document}
\label{firstpage}
\pagerange{\pageref{firstpage}--\pageref{lastpage}}
\maketitle

\begin{abstract}
The mass flows and energy cycles within the inner regions of galaxies exert a powerful influence on the evolution of the galaxy population. The centre of the Milky Way is the only galactic nucleus for which it is possible to resolve the physical mechanisms that drive these cycles, namely star formation and feedback, while also tracing global ($\gtrsim$100\,pc) processes which determine where and when star formation and feedback occur. We present an overview of ACES, the `Atacama Large Millimeter/submillimeter Array (ALMA) CMZ Exploration Survey', a $\sim1.5^{\prime\prime}$ angular resolution, $0.2-3$\,\kms spectral resolution ALMA Band 3 ($85-102$\,GHz), survey of the `Central Molecular Zone' (CMZ) -- the inner-100\,pc of the Galaxy (l = 359.4$^\circ$ to 0.8$^\circ$). ACES spectral setup is tuned to observe optimal tracers of the physical, chemical, and kinematic conditions in over 70 spectral features (e.g.\ HCO$^+$, HNCO, SiO, H40$\alpha$, complex molecules) of the gas in the CMZ, to derive the properties of all potentially star-forming Galactic Centre gas, from global scales (100\,pc) to dense $\sim$0.05\,pc structures that are expected to host individual star‑forming cores, down to sub-sonic ($<$0.4\,\kms) velocity resolution. In this overview paper, we provide the scientific justification for the ACES survey, explain the choice of observational setup, and describe the data legacy products. Finally, we show some of the initial ACES data which highlight the power of ACES' combination of high angular resolution, unprecedented spatial dynamic range, sensitivity, spectral resolution and spectral bandwidth as an illustration of how ACES aims to understand how global processes set the location, intensity, and timescales for star formation and feedback in the CMZ.
\end{abstract}

\begin{keywords}

Galaxy:centre --- ISM: kinematics and dynamics --- galaxies: star formation --- surveys --- ISM:molecules --- ISM:evolution
\end{keywords}


\section{Introduction}
\label{sec:intro}

\begin{figure*}
    \centering
    \includegraphics[width=1.0\textwidth]{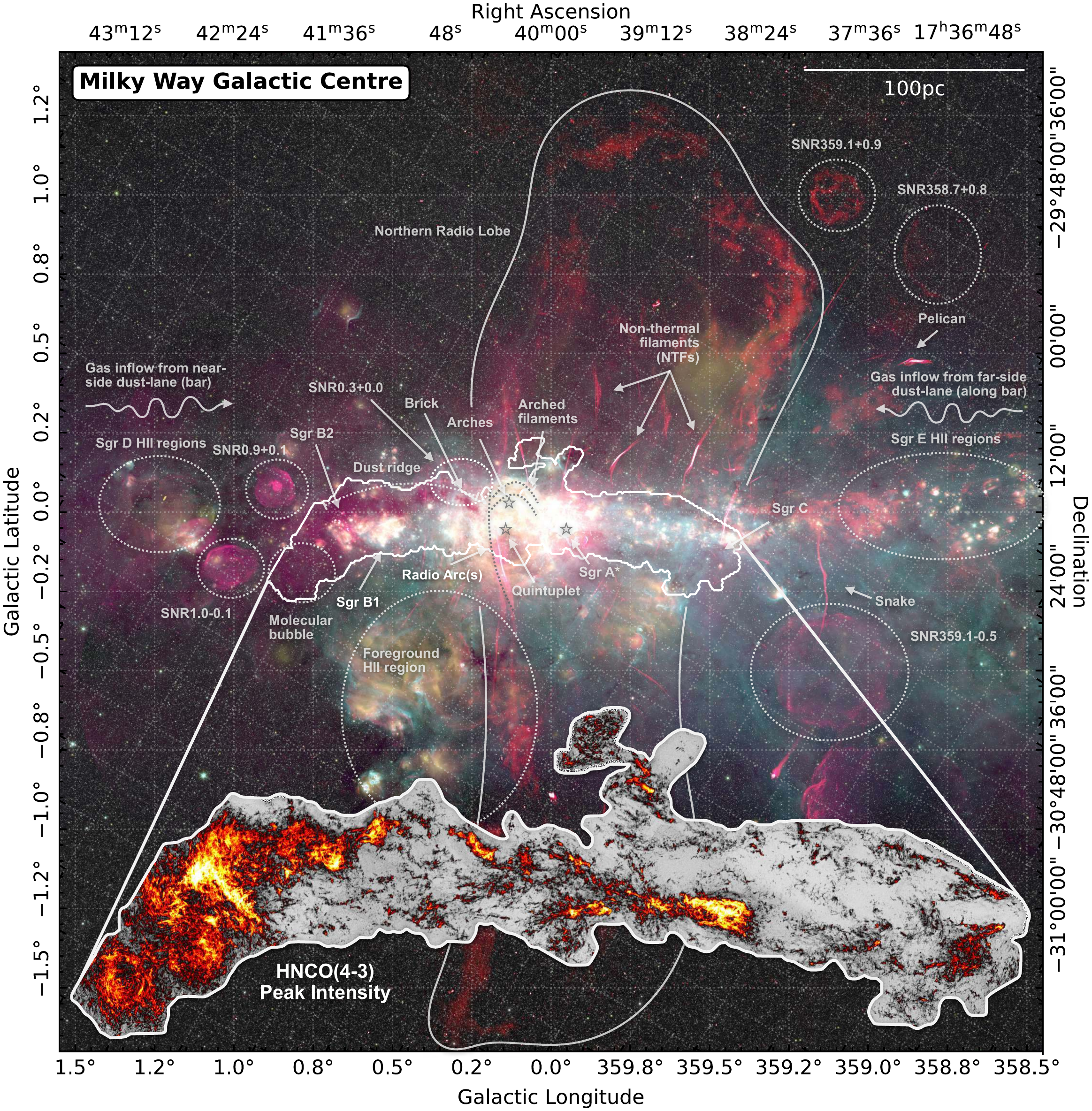} \\ \vspace{-1mm}
    \caption{Finding chart for the Galactic Center region. A colour composite of the 4.5\,$\mu$m (white) and 8\,$\mu$m (green) emission from the Spitzer GLIMPSE survey \citep{Churchwell2009}, 24\,$\mu$m (yellow) emission from the Spitzer MIPSGAL survey \citep{Carey2009}, and 20\,cm (red) emission observed by MeerKAT \citep{Heywood2019, Heywood2022} and the Green Bank Telescope (GBT; \citealp{Law2008}).  Overlaid are labels highlighting several features of interest across the Galactic Centre, including the central few 100\,pc known as the Central Molecular Zone (CMZ). Overlaid as a white contour is the coverage of the ACES survey (see Figure \ref{fig:ACES_coverage}). The inset zoom-in shows the ACES HNCO(1-0) peak intensity map. Figure~\ref{fig:ACES_coverage} shows the coordinates for the inset image. The background image for this Figure is adapted from \citealp{Henshaw2023}. An interactive version of the figure is available on the project homepage, https://sites.google.com/view/aces-cmz/home.}
    \label{fig:ACES_rgb}
\end{figure*}

\begin{figure*}
    \centering
    \includegraphics[trim=0 10mm 0 0, clip, width=1.0\textwidth]{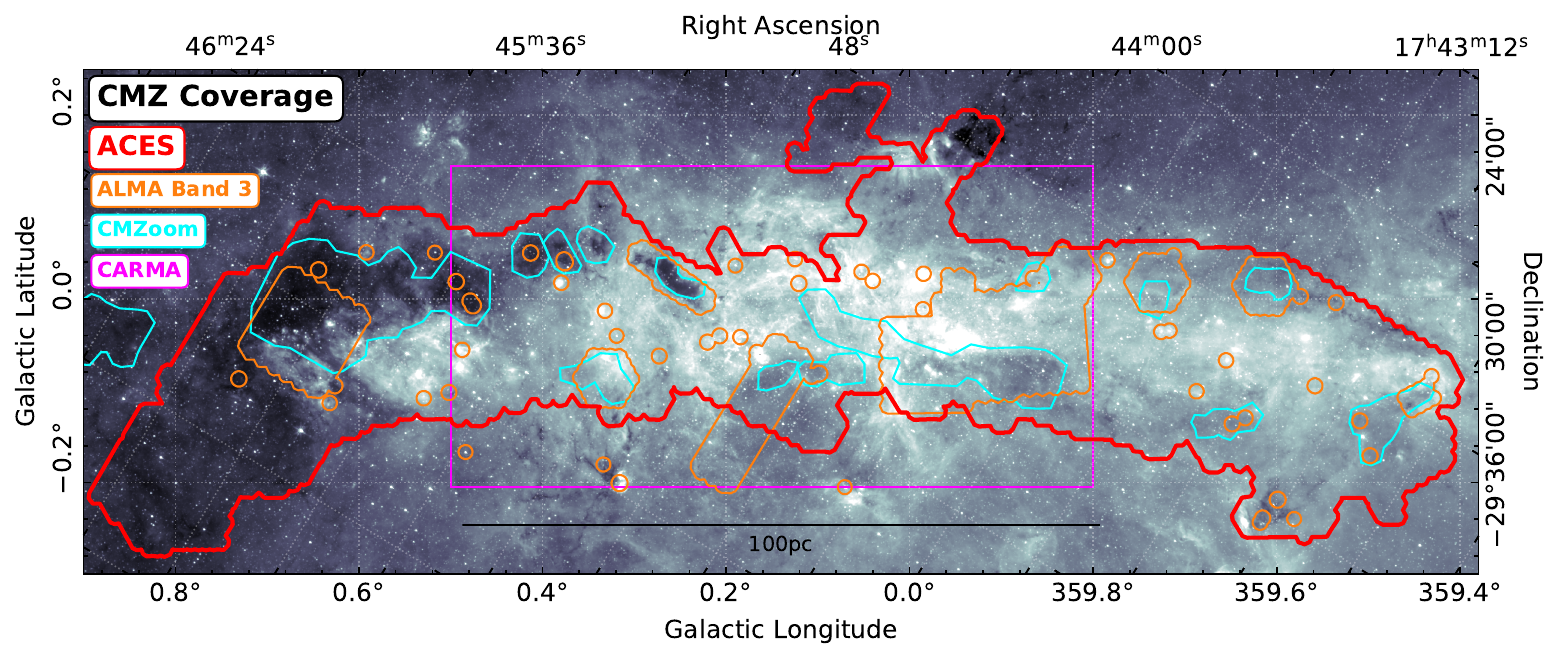} \\ \vspace{-3mm}
    \includegraphics[trim=0 0 0 11mm, clip, width=1.0\textwidth]{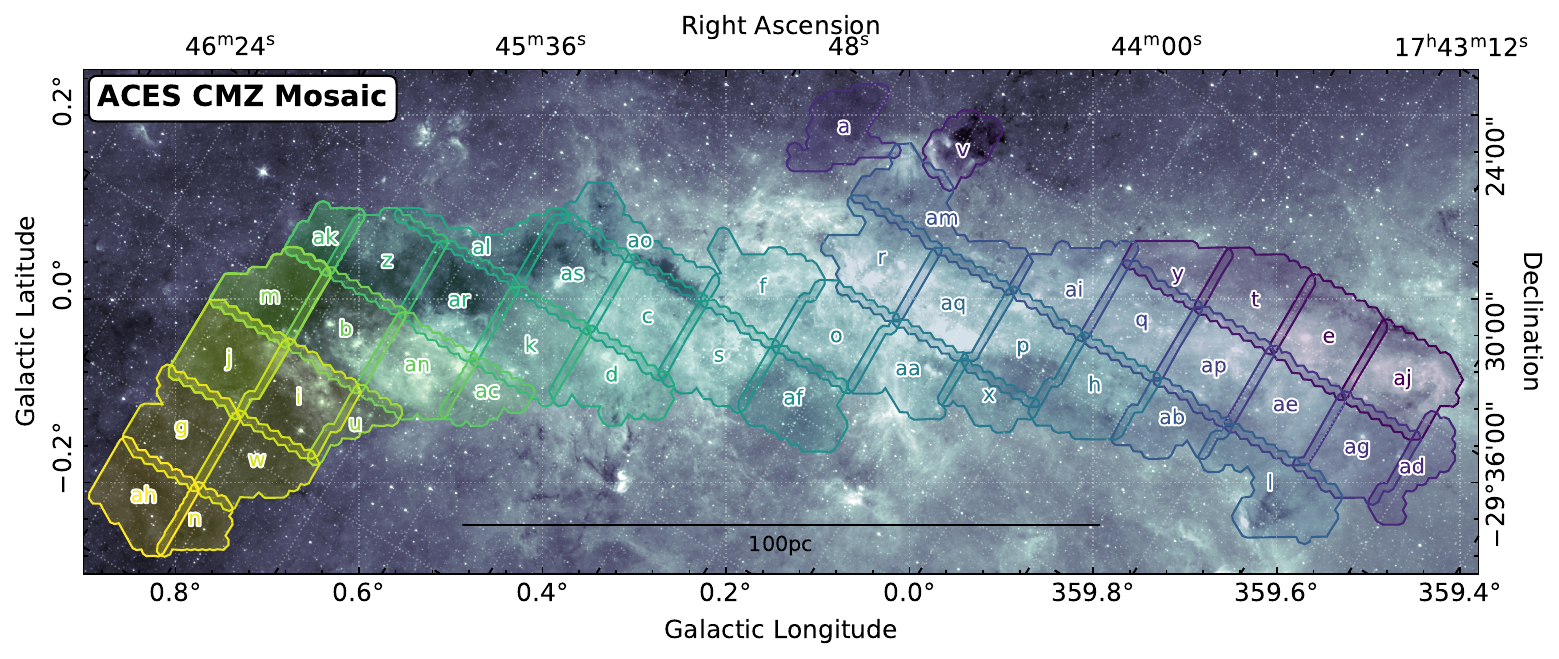} \vspace{-5mm}
    \caption{Coverage of the ACES survey. The upper panel displays the spatial coverage of the ACES survey, indicated by the red contour (see also Figure \ref{fig:ACES_rgb}). Overlaid orange contours denote the pre-ACES archival observations available in ALMA Band~3. The outline of the SMA \(1\,\mathrm{mm}\) CMZoom survey \citep{Battersby2020, Hatchfield2020_cmzoom, Callanan2023} is shown as the cyan contour, while the magenta rectangle represents the coverage of the CARMA survey \citep{pound2018}. The lower panel presents the individual mosaics comprising the ACES survey, each labelled with an alphabetical identifier. Both panels are superimposed on the \textit{Spitzer} \(8\,\mu\mathrm{m}\) image from the GLIMPSE survey \citep{Churchwell2009}.
    }
    \label{fig:ACES_coverage}
\end{figure*}

\begin{figure*}
    \centering
    \includegraphics[trim=6.5cm 7.7cm 6cm 6.5cm, clip, width=1.0\textwidth]{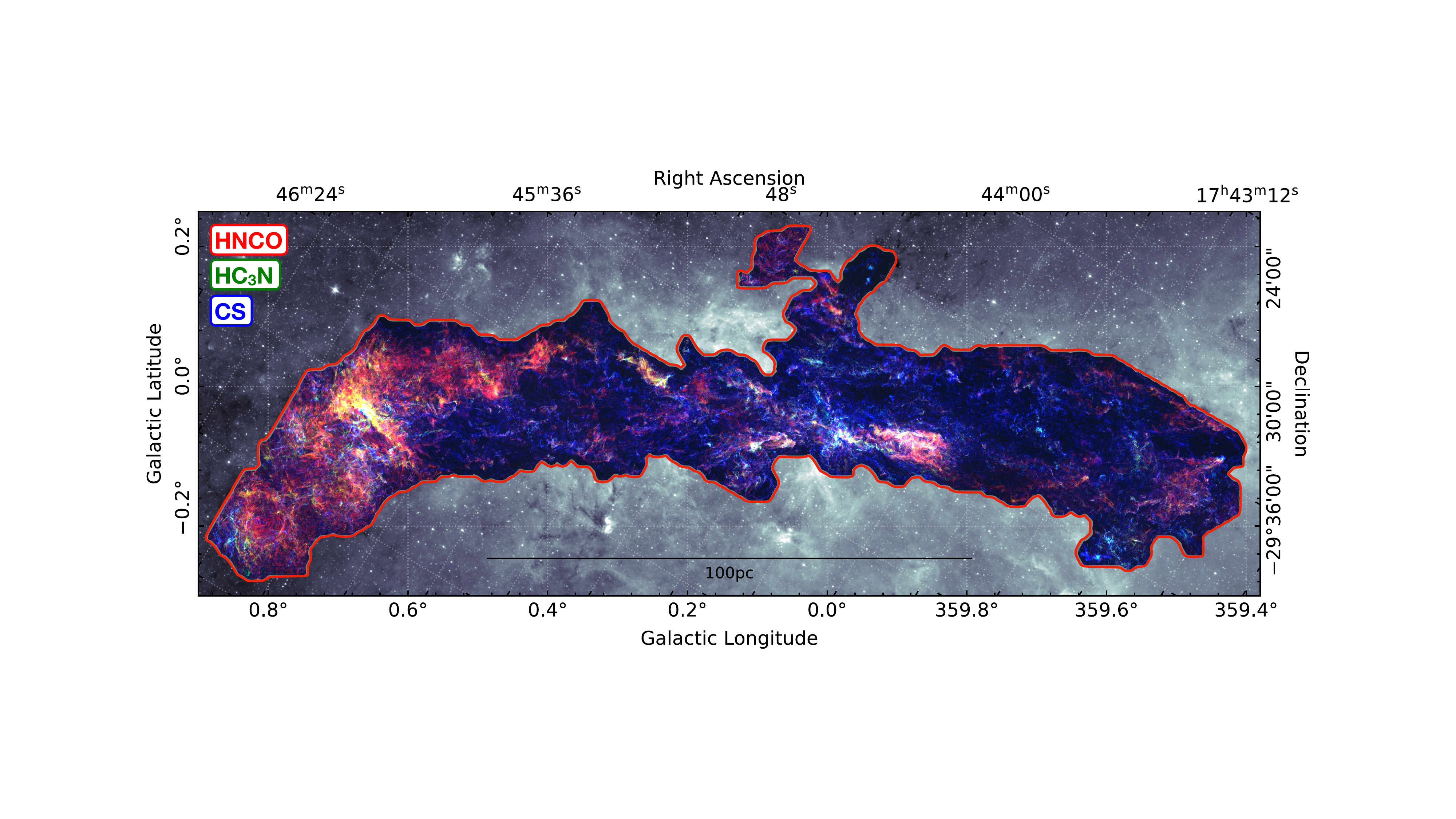} \\ \vspace{-8.5mm}
    \includegraphics[trim=6.5cm 7.2cm 6cm 9.8cm, clip, width=1.0\textwidth]{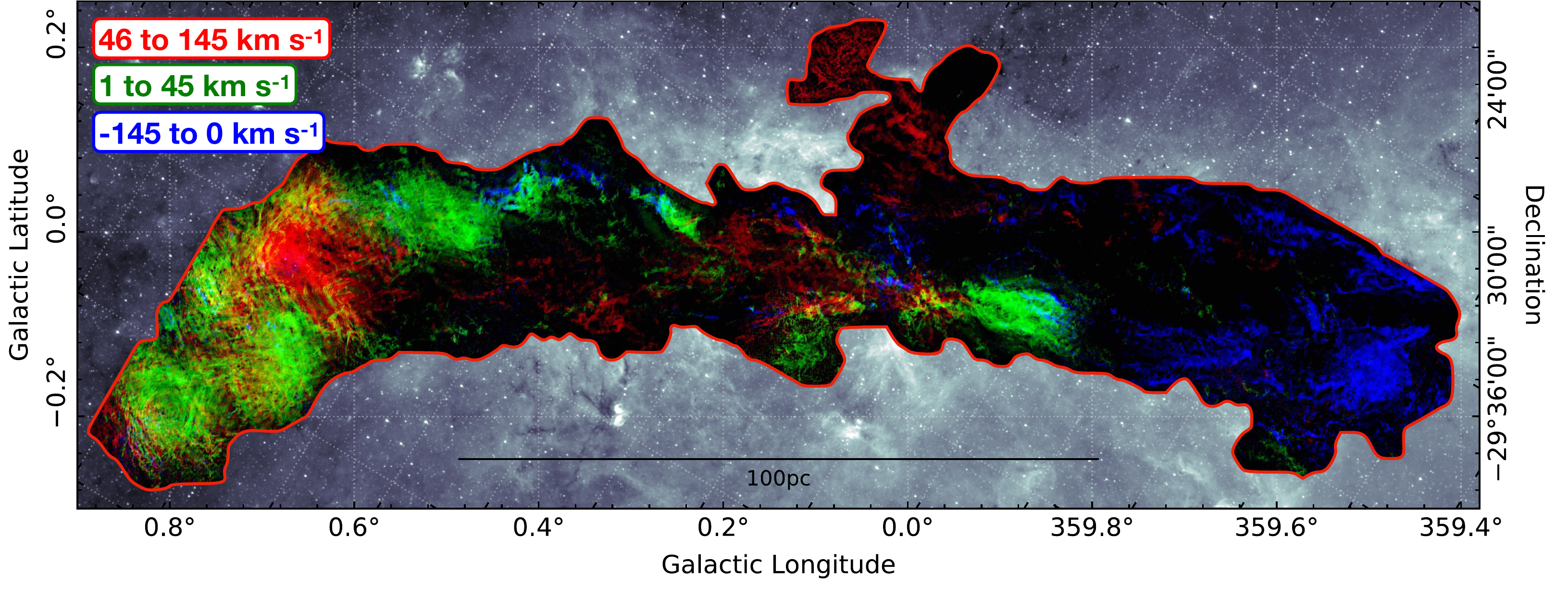}
    \\ \vspace{-2mm}
    \caption{Three-colour images of ACES observations. The upper panel shows an image of the ACES maximum intensity maps for CS $2-1$ (blue), HNCO $4-3$ (red), and HC$_{3}$N $11-10$ (green). The lower panel shows an image of a coarse rendering of the velocity structure in the ACES HNCO data, made by integrating the HNCO emission across three velocity ranges: 46 -- 145 \kms (red), 1 -- 45 \kms (green), and -145 -- 0 \kms (blue). Both panels are superimposed on the \textit{Spitzer} \(8\,\mu\mathrm{m}\) image from the GLIMPSE survey \citep{Churchwell2009}, and the red contour outlines the ACES coverage shown in Fig.\,\ref{fig:ACES_coverage} (see upper panel).}
    \label{fig:ACES_mom0_vel}
\end{figure*}

The mass flows and energy cycles in galactic nuclei play a key role in shaping the evolution of the galaxy population \citep[e.g.,][]{veilleux2005,Cicone2014,HeckmanBest2014,Harrison2017}. Bars and other large-scale stellar structures, as well as molecular clouds and star-formation complexes, evolve from these flows and cycles, both baryonic and radiative, and from historical gravitational processes including galaxy mergers. The central regions of galaxies are thus key to understanding these multiple cycles, evolutionary histories, and to quantifying how these global processes determine the location, intensity, and timescales for star formation and feedback \citep{Henshaw2023}. It is clear that many processes are involved -- a deep gravitational potential well, rapid differential rotation, turbulence injection, shocks, strong magnetic fields, etc. \citep{Kruijssen2014,Krumholz2015,Krumholz2017,Federrath2016,Sormani2019,Sormani2020,Hatchfield2021} -- the effects of which cascade from global (100\,pc), to cloud (few to 10\,pc), to dense  core-forming (0.05\,pc) scales \citep{Longmore2013b,Walker2015, Walker2018,Henshaw2016a,Kruijssen2019, Battersby2025a}. The centre of our Galaxy is the only nucleus for which we can resolve the physics of star formation and feedback down to the scale of individual protostellar cores \citep[e.g.][]{Ginsburg2018,Lu2020,Lu2021,Walker2021,Zhang2025,Xu2025}, thereby providing the crucial link between galactic-scale processes and the small-scale physics that ultimately regulate star formation and feedback \citep{Molinari2011, Longmore2013, Kruijssen2014, Henshaw2023}.

Determining the relative importance of these processes as a function of spatial scale, time and location requires measuring the physical and kinematic properties of gas contiguously over all these scales and comparing the gas properties with the comparatively well-established distribution of young stars and other feedback sources (Figure \ref{fig:ACES_rgb}). Previous single-dish observations have inadequate spatial resolution to achieve this goal \citep[e.g.,][]{Jones2012, Walsh2011, Purcell2012, Ginsburg2016, Krieger2017}, and existing interferometric studies at the targeted resolution are restricted to individual clouds \citep[e.g.,][see Figure \ref{fig:ACES_coverage}]{Battersby2020}. This discontinuity in spatial dynamic range leaves us with a fundamental gap in our knowledge between the small-scale physics of star formation and the global ($\sim$100\,pc) processes that control it. 

The goal of the ALMA Central Molecular Zone (CMZ) Exploration Survey (ACES) is to move beyond comparing a small number of the highest-density clouds observed with heterogeneous datasets, and to obtain a contiguous, uniform view of the dense gas, kinematics, and chemistry across the entire inner $\sim$100\,pc. By measuring the same set of physical quantities — mass surface density, velocity field, turbulent energy density, and chemical state — from cloud-complex scales down to 0.05\,pc structures containing individual proto-stellar cores, ACES is designed to connect global gas flows, local star formation, and feedback within a single, self-consistent framework.

Previous observational and theoretical work has established a basic picture of the mass flows within the CMZ. The Galactic bar transports gas inwards from the inner few kiloparsecs, where it accumulates in a ring or stream‑like structure at radii of $\sim$100\,pc with a total molecular gas mass of a few $10^7$\,M$_\odot$ and characteristic densities $\sim10^4$\,cm$^{-3}$ \citep[e.g.][]{Morris1996,Longmore2013, Kruijssen2014, Henshaw2023}. The gas is warm (T$\gtrsim$50–100\,K), highly turbulent (line widths $\gtrsim$ 10–20\,\kms on cloud scales), and pervaded by strong magnetic fields and an enhanced cosmic‑ray flux compared to the Galactic disc. On these global scales, orbital models and hydrodynamical simulations reproduce many features of the observed longitude–velocity structure and show how bar‑driven inflows, gravitational instabilities, and stellar feedback together set up a time‑dependent cycle of inflow, star formation, and outflow in galactic nuclei \citep[e.g.][]{Krumholz2015,Sormani2020, Tress2020}.

At the same time, the CMZ forms stars far less efficiently than expected from its dense gas content. Empirical star‑formation relations calibrated in the Galactic disc predict a star‑formation rate of order $0.5-1$\,M$_\odot$\,yr$^{-1}$ for the observed gas mass and density, but measurements based on ionising photon counts and young massive clusters instead find a roughly constant rate of $\sim$0.1\,M$_\odot$\,yr$^{-1}$ over the last $\sim5-10$\,Myr \citep[][]{Longmore2013, Barnes2017, Henshaw2023}. This mismatch implies either that star formation in the CMZ is episodic on at least a few Myr timescales, or that the local star‑formation law and the density threshold for star formation are strongly modified by the extreme environment — or both. High‑resolution observations of individual clouds support the idea that environmental effects are important: they reveal large internal velocity dispersions, strong shear and tidal forces, and dense, gravitationally bound cores that sometimes show very little current star formation \citep[][]{longmore2012, Longmore2013b, Rathborne2014, Ginsburg2018, Lu2020}.

Despite this progress, several key questions remain. Firstly, we still lack a quantitative, spatially resolved inventory of how mass, momentum, and energy are transported through the CMZ: how much is supplied by bar‑driven inflow, how much is redistributed by shear, turbulence, cloud–cloud collisions, and feedback, and how these processes vary with position along the orbit and with spatial scale. Secondly, the three‑dimensional geometry of the CMZ remains uncertain. Different models place well‑studied clouds on different orbits and at different distances from Sgr A, which directly affects interpretations of their tidal histories, exposure to radiation and cosmic rays, and their role in feeding the central black hole. Thirdly, we do not yet know whether there are preferred locations or dynamical phases at which CMZ clouds transition from quiescent to star‑forming, for example, at the contact points with the bar dust‑lanes, or downstream from pericentre passages where tidal compression is strongest. Finally, we lack a systematic measurement of the dimensionless parameters that appear in modern star formation theories \citep[e.g.\ virial parameter, Mach number, and turbulence driving parameter; see][]{FederrathKlessen2012} for all potentially star‑forming gas in the CMZ, and so we cannot yet test the extent to which the same physical prescriptions that describe star formation in the Galactic disc also apply in this extreme environment.

In this ACES overview paper, we explain how ACES is designed to address these open questions by providing a uniform, high‑dynamic‑range data set that connects global CMZ gas flows down to the $\sim$0.05\,pc dense structures that host star‑forming cores. In Section~\ref{sec:obs_design}, we explain the observational strategy, including the choice of tracers, spectral setups, and angular resolution that together provide the means to connect the large-scale mass flows with the formation of dense, star-forming cores. In Section~\ref{sec:simulation_network} we describe the complementary unified simulations framework developed to aid the interpretation of the data. Section~\ref{sec:science_objectives}~and~Appendix~\ref{sec:data_products} describe the science objectives and the open science approach, respectively. Finally, in Section~\ref{sec:data_highlights} we show some of the initial ACES data which highlight the power of ACES' combination of high angular resolution, unprecedented spatial dynamic range, sensitivity, spectral resolution and spectral bandwidth as an illustration of how ACES aims to understand how global processes set the location, intensity, and timescales for star formation and feedback in the CMZ.

In a series of companion papers, we  present the ACES continuum data \citep[Paper II;][]{Ginsburg2025_ACESII}, the high spectral resolution (0.2\,\kms) HNCO \& HCO$^+$ data \citep[Paper III;][]{Walker2025_ACESIII}, the two intermediate spectral resolution (1.7\,\kms) windows, containing SiO (2$-$1), SO (22$-$11), H$^{13}$CO$^+$ (1$-$0), H$^{13}$CN (1$-$0), HN$^{13}$C (1$-$0), and HC$^{15}$N (1$-$0) \citep[paper IV;][]{Lu2025_ACESIV}, and the two broad spectral windows containing CS(2$-$1), SO (23 $-$ 12), CH$_3$CHO 5(1,4) $-$ 4(1,3), HC$_3$N(11$-$10) and H40$\alpha$ \citep[Paper V;][]{Hsieh2025_ACESV}. 

\section{The ALMA CMZ Exploration Survey (ACES): Observational Design}
\label{sec:obs_design}

The Atacama Large Millimeter/submillimeter Array (ALMA) CMZ Exploration Survey (ACES) Large Program (Project code: 2021.1.00172.L; PI: S.~Longmore) provides uniform Band~3 coverage of the inner $1.5^\circ \times 0.5^\circ$ of the CMZ. ACES top-level observational goal was to achieve a uniform $\sim$1.5$\arcsec$ ($\sim$0.05\,pc) angular resolution, Band 3 ($85-102$\,GHz), molecular line and dust continuum survey of all gas in the inner 100\,pc of the Galaxy dense enough to form stars, at a line sensitivity of 1\,K ($\sim$32\, mJy/beam) per 0.2\,kms$^{-1}$ channel and a continuum sensitivity of 0.07\,mJy\,beam$^{-1}$, covering the full range of CMZ gas velocity (V$_{\rm LSR}\pm \sim$ 150\,\kms). In practice, due to the broad spectral windows used in most of the correlator settings, the velocity range for all lines other than HNCO and HCO$^+$ in the narrow spectral window extends far beyond V$_{\rm LSR}\pm \sim$ 150\,\kms\ in all spectral lines.

The definition of the ACES survey area was driven by the fundamental requirement to cover all gas in the inner $\sim$100 pc of the Galaxy expected to form stars. To meet this requirement, we used the H$_2$ column density map of the inner Galaxy constructed in the 3D CMZ project \citep{Battersby2025a}. Column densities and dust temperatures were derived by fitting a single‑temperature modified blackbody to the Herschel Hi‑GAL $70-500$\,$\mu$m images in each pixel \citep{Molinari2011}. The fits adopt a mean molecular weight $\mu = 2.8$, a gas‑to‑dust mass ratio of 100, and dust opacities based on the \citet{ossenkopf_henning1994} thin‑ice‑mantle coagulated grain model, parametrised as $\kappa_\nu  = 4.0$\,cm$^{2}$\,g$^{-1}$ at 505\,GHz with a fixed spectral index $\beta$ = 1.75. We then defined the ACES footprint on this map as the area with N(H$_2$) $\geq 1 \times 10^{22}$\,cm$^{-2}$ that are both within $359.4^\circ \leq l \leq 0.8^\circ$ and $|b| \leq 0.25^\circ$, and are kinematically associated with the well‑known $\sim$100\,pc radius molecular gas stream (Section 4.2). This threshold is identical to the one used by \citet{Battersby2025a} to characterise dense CMZ gas and matches the empirical column‑density threshold above which star formation becomes efficient in Galactic disc clouds \citep[e.g.][]{Lada12}. This choice therefore enables a direct comparison between star‑forming gas in the CMZ, the Milky Way disc and other galaxies.

The survey area imposed by the column density threshold of 10$^{22}$\,cm$^{-2}$ required making the largest contiguous map on the sky that ALMA has ever made. The survey area is broken up into 45 contiguous regions, following an alphabetised naming convention from `a' to `as', each containing up to 150 pointings per observation (Figure \ref{fig:ACES_coverage}).

The large survey area was a limiting factor in the choice of the spectral setup. At the time of the ACES observations, only Band 3 (observational wavelengths $\sim$3\,mm) had a sufficiently large primary beam to mosaic the survey area within the observational time constraints. The Band 3 spectral setup ($85-102$\,GHz) was selected to include transitions with a broad range of critical densities and excitation energies (see, e.g., Table~\ref{tab:aces_lineprops}), to trace the gas kinematic, physical, and chemical properties from global (tens of pc), to cloud (few pc), to dense structures on 0.05\,pc scales that host star‑forming cores (see Table 1 in \citet{Ginsburg2025_ACESII} for an overview of the spectral setup, and Table~\ref{table-simplemolecules} for a list of spectral lines in the observed frequency ranges). In practice, the effective excitation densities of these lines depend on radiative trapping, optical depth, and the elevated gas temperatures and line widths in the CMZ. As a result, the effective excitation densities can be orders of magnitude lower than simple critical density estimates \citep[][]{shirley2015, pety2017, kauffmann2017}. The ACES tracer selection therefore exploits their wide range in effective excitation conditions rather than assuming that any single transition maps onto a narrow density interval.

Two high spectral resolution (0.2\,\kms) windows centred on the reliable gas kinematic tracers HCO$^{+}$ and HNCO \citep{Henshaw2016a, He2021} were chosen to resolve the thermal linewidth \citep[0.4\,\kms\ in the 60\,K CMZ gas;][]{Ginsburg2016, Krieger2017} in global-to-cloud and cloud-to-core scale gas, respectively. Due to the narrow bandwidth in these high spectral resolution windows, the central frequency needed to be shifted as a function of spatial position to make sure the corresponding V$_{\rm LSR}$ range encompassed the CMZ gas velocities (between $\pm$150\,\kms; see Table A2 in \citet{Walker2025_ACESIII} to see the equivalent velocity shift for each of the 45 fields).

The other four spectral windows (SPWs) cover observed frequencies 85.96--86.43 GHz (SPW 25, resolution of 1.7 km s$^{-1}$), 86.67--87.13 GHz (SPW 27, resolution of 1.7 km s$^{-1}$), 97.66--99.54 GHz (SPW 33, resolution of $\sim$3 km s$^{-1}$), and 99.56--101.44 GHz (SPW 35, resolution of $\sim$3 km s$^{-1}$). 

The two intermediate spectral resolution (1.7\,km\,s$^{-1}$) windows are centred on high‑dipole molecules (e.g. HCO$^+$, HCN), classic shock tracers (e.g. SiO, HNCO) and their isotopologues. They were chosen to trace gas kinematics in intermediate‑ to high‑density gas and to help break degeneracies in opacity and excitation. Throughout this paper we refer to these high‑dipole species as “dense‑gas tracers” in the traditional extragalactic sense, while recognising that in nearby clouds their low-J emission is also excited in more diffuse gas \citep[e.g.,][see $\S$5.1.1]{pety2017, kauffmann2017, tafalla2023}. The $^{13}$C, $^{18}$O, and $^{15}$N isotopic substitutions of HCO$^{+}$, HCN, and HNC trace stellar nucleosynthesis gas enrichment \citep{Riquelme2010} and the properties of increasingly higher density (smaller scale) gas when the main isotope transitions become optically thick. Other lines were chosen as they have been found to reliably trace specific physical processes and regimes -- such as shocks (SiO), cosmic ray irradiation and star formation activity (e.g., HC$_{3}$N, HC$_{5}$N), ionised gas (H40$\alpha$, H50$\beta$) and chemical complexity (complex organic molecules) -- at least in Galactic star formation regions\footnote{We caution that due to the different conditions in the CMZ, reliable tracers of phenomena (e.g. shocks, hot-cores) in nearby star formation regions may not reliably trace the same phenomena in the CMZ.}. 

Finally, two broad bands were selected with 2.9\,\kms\ resolution, sufficient to trace the bulk kinematics and chemical complexity while allowing removal of lines to obtain continuum measurements. The bandwidths of the intermediate and broad spectral windows correspond to velocities of many thousands of \kms, easily sufficient to encompasses all the CMZ gas velocities.

\begin{table*}
\caption{Overview of the ACES spectral configuration. Shown for each 12m spectral window (SPW) ID are the lower ($\nu_{\rm L}$) and upper ($\nu_{\rm U}$) frequency bounds, native channel widths ($\Delta\nu$), and bandwidth. Prominent lines in each SPW are also shown.}
\label{tab:spectral_setup}
\begin{tabular}{cccccc}
\hline
SPW & $\nu_{\rm L}$ & $\nu_{\rm U}$ & $\Delta\nu$ & Bandwidth & Prominent Lines \\\hline
& $\mathrm{GHz}$ & $\mathrm{GHz}$ & $\mathrm{km~s^{-1}}$ & $\mathrm{GHz}$ &  \\\hline
25 & 85.9656 & 86.4344 & 0.849 & 0.46875 & HC$^{15}$N 1-0, SO 2$_2$-1$_1$, SiO 2-1 v=1 maser, H$^{13}$CN 1-0\\ 
27 & 86.6656 & 87.1344 & 0.842 & 0.46875 & H$^{13}$CO+ 1-0, SiO 2-1, HN$^{13}$C 1-0\\ 
29 & 89.1592 & 89.2178 & 0.103 & 0.05859 & HCO$^{+}$ 1-0\\ 
31 & 87.8959 & 87.9545 & 0.104 & 0.05859 & HNCO 4-3\\ 
33 & 97.6625 & 99.5375 & 1.485 & 1.875 & CS 2-1, CH$_3$CHO 5(1,4)–4(1,3) A–, H40$\alpha$, SO 2$_3$-1$_2$\\ 
35 & 99.5625 & 101.438 & 1.457 & 1.875 & HC$_3$N 11-10\\ 
\hline
\end{tabular}
\end{table*}

\begin{table*}
\begin{center}
\caption{\label{table-simplemolecules} Simple molecules included in the ACES setup. The transitions, rest frequencies and corresponding spectral windows (SPW) are given.}
\begin{tabular}{lcccc}
\hline
    Molecule & Transition & Rest Frequency(GHz) & 12m SPW \\
\hline
HC$^{15}$N & 1--0 & 86.0549 & 25 \\
SO & 2(2)--1(1) & 86.0939 & 25   \\
CCS & 7(6)--6(5) & 86.1813 & 25  \\
SiO v=1 maser& 2--1 & 86.2434 & 25 \\
H$^{13}$CN & 1--0 & 86.33992 & 25 \\
HCO & 1(0,1,2,1)--1(0,0,1,0) & 86.7083 & 27 \\
H$^{13}$CO$^{+}$  & 1--0 & 86.7543 & 27  \\
HCO & 1(0,1,1,1)--1(0,0,1,1) & 86.7774 & 27  \\
HCO & 1(0,1,1,0)--1(0,0,1,1) & 86.8057 & 27 \\
SiO v=0 & 2--1 & 86.8469 & 27  \\
HC$^{17}$O$^{+}$ & 1--0 & 87.0575 & 27   \\
HN$^{13}$C & 1--0 & 87.0908 & 27   \\
HNCO & 4--3 & 87.9252 & 31 \\
HCO$^{+}$ & 1--0 & 89.1885 & 29  \\
$^{34}$SO & 2(3)--1(2) & 97.7153 & 33 \\
CS & 2--1 & 97.9809 & 33   \\
$^{33}$SO & 2(3,4)--1(2,3) & 98.4892 & 33   \\
$^{33}$SO & 2(3,5)--1(2,4) & 98.4936 & 33   \\
HC$_{5}$N &  37--36& 98.5125 & 33 &  \\
C$_{3}$N & 10(11)--9(10) & 98.940 & 33 \\
C$_{3}$N & 10(10)--9(9) & 98.958 & 33 \\
H40$\alpha$ &  & 99.02295 & 33   \\
H50$\beta$ & & 99.22521 & 33   \\
SO & 2(3)--1(2) & 99.2999 & 33  \\
HC$^{13}$CCN & 11--10 & 99.6518 & 35  \\
HCC$^{13}$CN & 11--10 & 99.6615 & 35 \\
H$_{2}$C$^{34}$S & 3(1,3)--2(1,2) & 99.7740 & 35 \\
CCS & 8(7)--7(6) & 99.8665 & 35 \\
S$^{18}$O & 3(2)--2(1) & 99.8038 & 35  \\
SO & 5(4)--4(4) & 100.0296 & 35 \\
HC$_{3}$N & 11--10 & 100.0764 & 35   \\
HC$_{5}$N &  38--37& 101.1747 & 33 &  \\
NS$^{+}$ & 2--1& 100.1985 & 35   \\
HC$_{3}$N, v6=1 &11(1)--10(-1)  & 100.2406 & 35 &  \\
H$_{2}$C$^{34}$S & 3(0,3)--2(0,2) & 101.2843 & 35 \\
HC$_{3}$N, v6=1 &11(-1)--10(1)  & 100.3194 & 35 &  \\
HC$_{3}$N, v7=1 &11(1)--10(-1)  & 100.3224 & 35 \\
HC$_{3}$N, v7=1 & 11(-1)--10(+1)  & 100.46617 & 35   \\
CH$_3$NC & 5(1)--4(1) & 100.5242 & 35 \\
CH$_3$NC & 5(0)--4(0) & 100.5265 & 35 \\
HC$_{3}$N, v7=2 & 11(0)--10(0)  & 100.70878 & 35  \\
HC$_{3}$N, v7=2 & 11(2)--10(-2)  & 100.71106 & 35  \\
HC$_{3}$N, v7=2 & 11(-2)--10(2)  & 100.71439 & 35  \\
H$_{2}$C$^{34}$S & 3(0,3)--2(0,2)&  101.2843& 35   \\
H$_{2}$CO & 6(1,5)--6(1,6) & 101.332987 & 35 \\
\hline
  \normalsize
   \end{tabular}
   \end{center}
\end{table*}

\begin{table}
\centering
\caption{Excitation properties for the main ACES diagnostic transitions. We list the rest frequency $\nu_0$, upper-level energy $E_u/k$, and the formal critical density $n_{\rm crit}$ evaluated at $T=100$~K, defined as $n_{\rm crit}=A_{ul}/\sum_{l<u}\gamma_{ul}$ using LAMDA collisional rate coefficients (para-H$_2$ when available). For HN$^{13}$C, $n_{\rm crit}$ is estimated from HNC collision rates with $A_{ul}$ scaled as $\nu^3$.}
\label{tab:aces_lineprops}
\begin{tabular}{l l r r r}
\hline
Molecule & Transition & $\nu_0$ (GHz) & $E_u/k$ (K) & $n_{\rm crit}$ (cm$^{-3}$) \\
\hline
HC$^{15}$N          & $J=1-0$                  &  86.0550 &  4.1 & $3.0\times10^{6}$ \\
SO                 & $2_{2}-1_{1}$             &  86.0939 & 19.3 & $3.0\times10^{4}$ \\
H$^{13}$CN         & $J=1-0$                  &  86.3399 &  4.1 & $3.0\times10^{6}$ \\
H$^{13}$CO$^{+}$   & $J=1-0$                  &  86.7543 &  4.2 & $2.1\times10^{5}$ \\
SiO ($v=0$)        & $J=2-1$                  &  86.8470 &  6.3 & $1.3\times10^{5}$ \\
HN$^{13}$C         & $J=1-0$                  &  87.0908 &  4.2 & $3.7\times10^{5}$ \\
HNCO               & $4_{0,4}-3_{0,3}$         &  87.9252 & 10.6 & $4.0\times10^{4}$ \\
HCO$^{+}$          & $J=1-0$                  &  89.1885 &  4.3 & $2.1\times10^{5}$ \\
CS                 & $J=2-1$                  &  97.9810 &  7.1 & $1.7\times10^{5}$ \\
SO                 & $2_{3}-1_{2}$             &  99.2999 &  9.2 & $1.8\times10^{5}$ \\
HC$_3$N            & $J=11-10$                 & 100.0764 & 28.8 & $1.5\times10^{5}$ \\
\hline
\end{tabular}
\end{table}

ACES' targeted angular resolution of 1.5$\arcsec$ (0.05\,pc) was chosen to match the expected size of dense gas hosting (pre)star-forming cores in CMZ clouds \citep{Ginsburg2018}. As described in detail in Paper II, the angular resolution of the delivered observations largely fell within ALMA's 20\% tolerance level of the requested angular resolution. Additional 7m Atacama Compact Array (ACA) and total power (TP) observations were conducted to recover emission from large spatial scales for the line data, as existing single-dish data do not cover the full ACES spectral window and suffer from artifacts in large spatial and frequency ranges \citep{Jones2012}. Single-dish data from the Mustang Galactic Plane Survey \citep{Ginsburg2020} provide the 3mm continuum zero-spacing flux.

Based on the observed brightness temperatures in existing single-dish data \citep{Jones2012}, previous ALMA observations of CMZ clouds at a similar resolution and sensitivity, and numerical simulations with chemistry and radiative transfer modelling \citep{Petkova2023a}, ACES targeted a root-mean-square (RMS) line sensitivity of ${\rm RMS} =$ 1.0, 0.6 and 0.25\,K in 0.2, 1.7 and 3\,\kms\ channels, respectively, to detect all key lines necessary to achieve ACES' science goals at high significance in essentially all pointings. 

The corresponding continuum sensitivity is 0.07\,mJy\,beam$^{-1}$. To translate this into a gas column density, we assume that the emission is optically thin, adopt the \citet{ossenkopf_henning1994} thin‑ice‑mantle dust grain opacity law with $\beta = 1.75$, a gas‑to‑dust mass ratio of 100, and a mean molecular weight $\mu = 2.8$. For a $1.5''$ Gaussian beam, a 1$\sigma$ continuum sensitivity of 0.07\,mJy\,beam$^{-1}$ then corresponds to
\[
 N(\mathrm{H_2}) \simeq 1.2\times10^{22}
 \left(\frac{T_\mathrm{d}}{30~\mathrm{K}}\right)^{-1}
 \mathrm{cm}^{-2},
\]
i.e. of order $10^{22}$\,cm$^{-2}$ for dust temperatures typical of CMZ dense gas \citep[T$_\mathrm{d}$$ \sim 20-35$\,K,][]{Battersby2025a}. Systematic uncertainties in $\kappa_\nu$, gas‑to‑dust ratio, and dust temperature are at least at the factor‑of‑two level \citep{Battersby2025a}. Typical CMZ clouds peak at $\geq 10^{23}$\,cm$^{-2}$. The adopted footprint threshold N(H$_2$) $= 10^{22}$\,cm$^{-2}$ is consistent with, but not uniquely determined by, the continuum sensitivity, and is primarily motivated by the empirical star‑formation threshold discussed above.

Assuming the same dust properties, the continuum sensitivity of 0.07\,mJy\,beam$^{-1}$ is sufficient to detect $\sim10$\,M$_\odot$, 30\,K dense structures at 5$\sigma$. At $1.5''$ ($\sim0.05$\,pc) resolution, these structures are expected to contain one or more unresolved $\sim10^3$\,au star‑forming cores. ACES therefore provides a complete census of all dense, core‑bearing structures more massive than $\sim$10\,M$_\odot$ that can act as precursors of high‑mass stars in the CMZ.

Figure \ref{fig:ACES_mom0_vel} shows an example comparison of the dense gas distribution in different tracers and a velocity structure map from the ACES survey. Full details of the data reduction are given in the corresponding companion papers \citep[Paper II (continuum), Paper III (high spectral resolution), Paper IV (intermediate spectral resolution), Paper V (broad band) --][respectively]{Ginsburg2025_ACESII, Walker2025_ACESIII, Lu2025_ACESIV, Hsieh2025_ACESV}.

Table~\ref{tab:cmz_surveys} summarises the key observational parameters of ACES compared to previous millimetre surveys of the CMZ. Single‑dish studies such as the Mopra 3‑mm survey \citep{Jones2012} provide wide‑field ($2.5^\circ\times0.5^\circ$) coverage with excellent brightness sensitivity ($\sim 0.05$ K per $\sim 3.5$ km s$^{-1}$ channel) at $\sim 40''$ resolution, but cannot resolve the $\sim$0.05\,pc scales that are critical for connecting dense cores to global gas flows. Interferometric surveys like CARMA \citep{pound2018} and CMZoom \citep{Battersby2020, Hatchfield2020_cmzoom, Callanan2023} achieve $\sim 3\text{–}8''$ resolution, but either cover only the central $0.7^\circ\times0.4^\circ$ or sample only the highest column density gas. ACES extends high‑resolution mapping to essentially all gas dense enough to form stars in the inner $\sim$100\,pc, delivering $\sim1.5''$ (0.05\,pc) resolution over $1.5^\circ\times0.5^\circ$ with line rms of 1.0, 0.6 and 0.25\,K (0.2, 1.7 and 3.0\,km\,s$^{-1}$ channels) and a continuum rms of 0.07 mJy\,beam$^{-1}$, corresponding to $\sim 4\times10^{-3}$\,K. When smoothed to the coarser beams of the single‑dish surveys, the ACES cubes achieve comparable brightness sensitivity while retaining the interferometric information needed to link cloud‑ and core‑scale structure to the global CMZ environment.

\begin{table*}
\centering
\caption{Comparison of ACES with previous millimetre surveys of the Central Molecular Zone: the Mopra 3-mm CMZ survey \citep{Jones2012}, CMZoom \citep{Battersby2020}, and the CARMA 3-mm CMZ survey \citep{pound2018}. RMS values are typical $1\sigma$ sensitivities in brightness temperature. The
channel width $\Delta v$ is given for the line data.}
\label{tab:cmz_surveys}
\begin{tabular}{lcccccc}
\hline
Survey & Facility & $\nu$ & Area & $\theta_{\rm res}$ & Line rms & Continuum rms \\
       &          & (GHz) & (deg$^2$) & (arcsec) & (K, $\Delta v$) & (K) \\
\hline
ACES & ALMA+ACA+TP & 85--102 & $1.5^\circ\times0.5^\circ$ & 1.5 &
1.0 (0.2 km s$^{-1}$); & $\sim 4\times10^{-3}$ \\
     &             &        &                 &     &
0.6 (1.7 km s$^{-1}$); & (0.07 mJy beam$^{-1}$) \\
     &             &        &                 &     &
0.25 (3.0 km s$^{-1}$) & \\
Mopra 3-mm CMZ & Mopra 22-m & 85--93 & $2.5^\circ\times0.5^\circ$ & 40 &
$\sim 0.05$ (3.5 km s$^{-1}$) & -- \\
CMZoom (1.3 mm) & SMA+APEX+BGPS & 230 & 0.10\,deg$^2$ & 3.2 &
$\sim 0.2$ (1.1 km s$^{-1}$) & $8\times10^{-3}$ \\
     &             &     &      &     & & (3.5 mJy beam$^{-1}$) \\
CARMA 3-mm CMZ & CARMA-15+8+Mopra & 86--98 & $0.7^\circ\times0.4^\circ$ &
$7.6\times3.5$ (cont.) & $\sim 0.5$ (2.5 km s$^{-1}$) &
$(1$--$2)\times10^{-2}$ \\
     &             &     &      &
$13\times6$ (lines) & & (2--4 mJy beam$^{-1}$) \\
\hline
\end{tabular}
\end{table*}

It is useful to place ACES alongside recent ALMA Large Programmes that target dense cores in the Galactic disc. ALMAGAL \citep{molinari2025, sanchez-monge2025} imaged 1013 dense clumps in Band 6 (230\,GHz) at an almost uniform spatial resolution of $\sim800-2000$\,au (median 1400\,au), resolving compact cores with typical diameters of $\sim10^3$\,au. The ALMA‑IMF survey \citep{motte2022, ginsburg2022} imaged 15 massive Galactic disk protoclusters at a physical resolution of a few $\times 10^3$\,au, with angular resolutions of $0.3''-1.5''$ over sources at distances of $2-5.5$\,kpc, to measure the core mass function and core kinematics in nearby high‑mass star‑forming regions. ACES achieves $\sim1.5''$ ($\sim0.05$\,pc $\approx 10^4$\,au) resolution over essentially all star‑forming gas in the inner $\sim100$\,pc. At this resolution, individual $\lesssim10^3$\,au cores are not resolved. Instead, ACES identifies and characterises the dense $\sim0.05$\,pc structures that host one or more star‑forming cores, together with their larger‑scale kinematic and chemical environment. ACES is therefore complementary to ALMAGAL and ALMA‑IMF, trading core‑scale resolution for contiguous coverage and a uniform view of core‑bearing structures in the extreme CMZ environment.

\section{The ALMA CMZ Exploration Survey (ACES): Unified Simulation Network}
\label{sec:simulation_network}

It was clear from the initial conception of the survey, that a complementary suite of numerical simulations able to self-consistently trace the baryon cycle as well as gas dynamics and star formation at the same spatial scales as the ACES observations were a crucial component of achieving the core science objectives. With appropriate post-processing to produce accurate synthetic images, the simulations aim to provide the necessary numerical framework to interpret the key physical agents (gravitational forces, rapid differential rotation, turbulence injection, shocks, strong magnetic fields, etc.,) controlling the evolution from global (100\,pc), to cloud (few to 10\,pc), to dense $\sim$0.05\,pc scale structures hosting individual star-forming cores.

\begin{figure*}
\centering\includegraphics[width=1.0\textwidth]{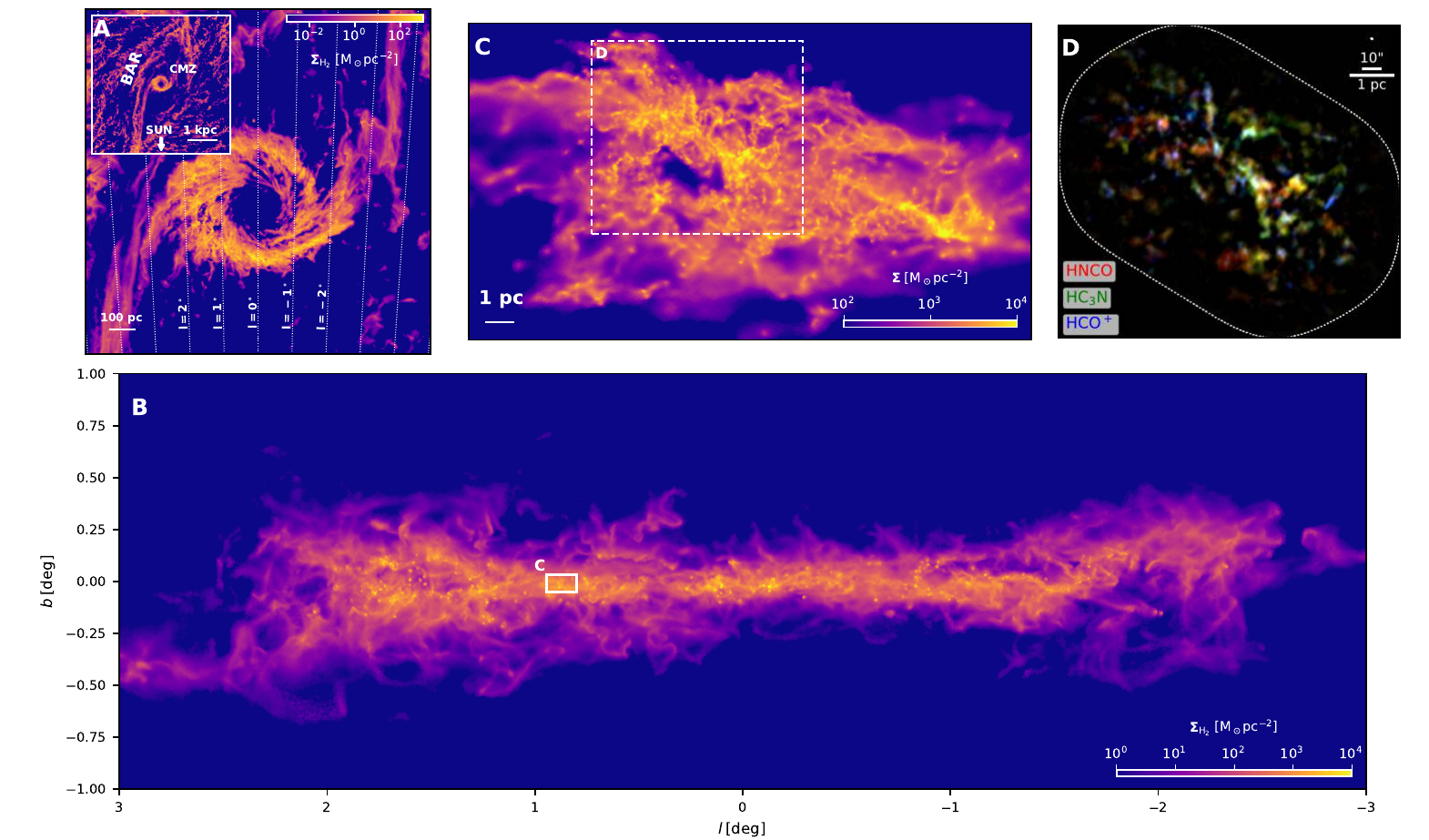}
\caption{Example of simulations in development that will help with interpretation of ACES data. White boxes illustrate zoom-in/-out regions in other panels. \emph{Panels A-B}: a top-down and plane-of-the-sky H$_2$ projection of a large-scale simulation of the gas flow in the Galactic bar with star formation, stellar feedback (supernovae, ionising radiation) and magnetic fields (ideal MHD). Simulations like this one can follow the formation of the CMZ and of individual molecular clouds self-consistently from kpc scales down to the cloud scale, but cannot resolve star formation.
\emph{Panel C:} simulation of an individual molecular cloud on a CMZ orbit through the Galactic potential, that can resolve star formation (Adapted from \citealt{Dale2019,Kruijssen2019}). \emph{Panel D:} a synthetic line emission image created by post-processing the simulation snapshot from Panel C (Adapted from \citealt{Petkova2023a}).}\label{fig:simulations}
\end{figure*}

Due to the large spatial dynamic range of the gas flow and star formation within the Galactic Centre, we have chosen to address the modelling through a multi-scale framework of numerical simulations. We use these simulations as numerical experiments in which we control the physical assumptions in order to constrain the effects and importance of various mechanisms (such as magnetic fields, stellar feedback, etc.). Specific simulations are then selected based on the science goals of each observational study, and are used for interpreting the observations. A recent application of this approach can be seen in \citet[][discussed in more detail below]{Pare2025}, where HNCO filaments were identified in the ACES data and compared to filaments in simulations including a Milky-Way-like gravitational potential  and magnetic fields \citep[similar to][]{Tress2024}. 
 
On the largest scales, we build models that include the entire barred region, and are therefore able to self-consistently follow the gas flow from the disk towards the Galactic Centre. These simulations build upon previous work where the gas evolves in a background Milky-Way-like gravitational potential \citep{Sormani2020, Tress2020, Hatchfield2021}. As part of the ACES project, we have expanded on these by updating the physical prescription including magnetic fields \citep{Tress2024} and radiative feedback \citep{Tress2025}. 
 
A detailed description of our modeling can be found in \citet{Tress2025}. This is a state-of-the-art simulation aimed at studying the interstellar medium (ISM) of the CMZ in the broader Milky Way context. This simulation uses the {\sc arepo} \citep{Springel2010} code, and the ISM is evolved in a Milky-Way-like background barred Galactic potential \citep{Hunter2024}. This creates a continuous flow of gas towards the Galactic Centre and results in the self-consistent formation of a CMZ-like region (see panels A and B in Figure \ref{fig:simulations}). We include non-equilibrium hydrogen and CO chemistry, magnetic fields, star formation through stochastic star particles, stellar feedback in the form of individual supernovae and radiation feedback by solving the radiative transfer equation on-the-fly with {\sc sweep} \citep{Peter2023}. Accretion onto the central supermassive black hole (SMBH) is followed as well. This simulation, and variations of it, will be used to study the time-evolution of the gas in the region as a whole, help interpret its complex 3D geometry, identify sites of increased star formation and understand the role of feedback.
 
Although this simulation includes many physical processes, it is unable to resolve the densest parts of the ISM. Therefore, some scientific goals may require simplified simulations to achieve the necessary resolution. For instance, in \citet{Pare2025} where the focus was on HNCO filaments and higher resolution simulations were needed, we excluded the effects of gas-self-gravity, star formation and feedback \citep[similar to][]{Tress2024}. This allowed us to understand how magnetic fields and Galactic shear shape filaments in the CMZ, and compare them to the observed filaments.
 
For cases where both higher resolution and advanced physics is needed, we are developing high-resolution zoom-in simulations (e.g.\ \citealt{Hatchfield2021}; Lipman et al., in prep., Feng at al., in prep.). These zoom-ins utilise the same {\sc arepo} setup as the large-scale simulations described above, and aim to incorporate as much key physics as possible (e.g.\ up-to-date Galactic potential, magnetic fields, star formation and feedback, etc), while evolving a small region at an increased resolution ($\lesssim 1$~M$_\odot$) for a short integration time ($\sim 10$~Myr). With these simulations we model the large-scale Galactic environment, while also reaching resolutions in the CMZ to resolve individual molecular clouds with star-forming cores (Lipman et al., in prep). We are also developing radius-dependent resolution simulations to follow the formation and evolution of the circum-nuclear disk (CND; $R\sim5\,\mathrm{pc}$) self-consistently from the larger-scale inflow (Feng et al., in prep.).
 
To complement the Galactic bar simulations introduced above, we also utilise existing and develop new simulations of individual clouds on CMZ orbits through the Galactic Centre gravitational potential \citep[e.g.][see panel C of Figure \ref{fig:simulations}]{Dale2019, Kruijssen2019}. Such simulations can resolve star formation and test the physical mechanisms that affect it on a small scale. They can also be set up to match properties of known clouds \citep[e.g.,][]{Clark13}, and attempt to reproduce their observed morphology, especially after being post-processed with radiative transfer (panel D of Figure \ref{fig:simulations}). However, these models do not include interactions between the clouds, and are more suitable for clouds that can be treated as individual entities. Some recent examples of post-processing and analysis of these simulations include \citet{Petkova2023a} and \citet{Petkova2023b}. The underlying simulations \citep{Dale2019, Kruijssen2019} were produced with the smoothed particle hydrodynamics code  {\sc gandalf} \citep{Hubber2018}, and included gas self-gravity, star formation and the external Galactic gravitational potential. The analysis studies were conducted in anticipation of the ACES dataset and included comparisons to a pre-existing ALMA dataset of similar spatial resolution \citep{Rathborne2014, Rathborne2015}, with \citet{Petkova2023b} being explicitly developed as part of the ACES project.
 
Section~\ref{sub:chem_phys_modelling} describes dedicated astrochemistry models that are being developed to interpret the plethora of lines detected with ACES. These high chemical fidelity models naturally complement the above hydro simulations, which contain numerous physical processes but only very simplified chemistry.

\section{ACES Science Objectives}
\label{sec:science_objectives}

Gas flows in the inner kpc of the Milky Way are typical of those in other barred-spiral galaxies \citep{Binney1991, RodriguezFernandez2008, Tress2020, Stuber2023}. Gas, funnelled radially inward from the disk, builds up a massive (few 10$^7$\,M$_{\odot}$), dense (10$^4$\,cm$^{-3}$) gas stream orbiting the centre \citep[$r \sim 100$\,pc;][]{Bally2010, Molinari2011, Kruijssen2015}. Empirical star formation relations relating the amount of dense gas to the star formation rate predict this gas should be producing $\sim$1\,M$_{\odot}$\,yr$^{-1}$ of stars \citep{Longmore2013}, yet the star formation rate has been constant at $\lesssim$0.1\,M$_{\odot}$\,yr$^{-1}$ to within a factor 2 for the last $\sim$5--10\,Myr \citep{Barnes2017, Elia2025}. This poses fundamental problems for star formation theories given that the free-fall time and turbulence dissipation timescale are both only $\sim$1\,Myr. Theory, numerical simulations, and pilot studies of individual clouds all show that global ($\lesssim$100\,pc) gas dynamics play a fundamental role in determining the star formation activity (or lack thereof) throughout the CMZ \citep{Kruijssen2014, Rathborne2015, Federrath2016, Sormani2020}. With observations of transitions covering the required critical densities and excitation energies (see, e.g., Table~\ref{tab:aces_lineprops}), ACES aims to provide the mass distribution and velocity structure at a contiguous spatial dynamic range of 200\,pc / 0.05\,pc $\approx 10^{3.5}$, down to the $\sim0.05$\,pc dense structures that host star‑forming cores. Combined with a suite of complementary numerical simulations, this crucial global-to-protostellar link overcomes the fundamental challenge of existing studies, making it possible to achieve ACES' core aim of quantifying how global ($\sim 100\, \text{pc}$) processes determine the location, intensity, and timescales for star formation and feedback. This core goal is split into four science objectives.

\subsection{Science Objective 1: Determine the mechanisms driving mass flows as a function of size scale and location}
\label{sub:sci_obj_1}

Comparing the observed mass distribution and kinematic structure with numerical simulations of gas evolving in the known gravitational potential, ACES aims to determine the relative influence of different physical processes (orbital motion, shear, turbulence, mass accretion, etc.) in shaping the physical structure and internal dynamics of the $\sim$0.05\,pc dense, core‑bearing structures within gas clouds, as a function of location across the CMZ \citep{Kruijssen2019, Dale2019, Sormani2020, Tress2020}. 

The first step to achieve this goal is to identify coherent velocity structures through multi-line spectral decomposition followed by multi-scale structure-finding decomposition of these velocity components. The 3\,mm dust continuum emission provides the mass per pixel along each line of sight. For lines of sight with a single velocity component, the mass-weighted centroid velocity, velocity dispersion, and velocity gradients along/across the streams can then be derived from optically thin tracers (e.g. H$^{13}$CO$^+$, H$^{13}$CN, HN$^{13}$C) in the higher spectral resolution windows. For lines of sight with more than one velocity component, the same technique can be used after making sensible decisions about how the mass along that line of sight is split between the different velocity components (e.g., weighting by the integrated intensity of optically thin lines). After using different geometric models to convert the position-position-velocity information to a plausible 3D gas structure \citep[see e.g.][]{Walker2025}, integrating $\rho v$ over surfaces perpendicular to the local flow direction will then provide estimates of mass and momentum fluxes along the bar-driven inflow and around the $\sim$100\,pc stream. On smaller scales, converging gradients and blue-skewed profiles in optically thick vs.~thin line pairs will make it possible to identify and quantify multi-scale accretion flows down to $\sim$0.05\,pc scales, within which individual stars are forming.

Combining ACES data with the temperature maps from the JACKS survey (Mills et al, in prep.) will make it possible to subtract both the thermal and instrumental contributions from the measured line widths. Combining these line widths with the dust-based mass density field will provide the non-thermal velocity dispersion and kinetic energy density as a function of physical scale.

Where SiO and SO emission and recombination lines delineate expanding shells or feedback-driven structures, expansion velocities measured from their line profiles together with shell masses from the continuum will provide momentum and energy injection rates, as already demonstrated for the M0.8$-$0.2 ring \citep[Section 5.2,][]{Nonhebel2024}.

These observationally derived fluxes and energy densities can then be compared to the suite of zoom-in and galaxy-scale simulations (Section 3) by running the same line-of-sight integration and line-fitting on synthetic ACES cubes, enabling a direct assessment of the roles of orbital motion, shear, turbulence, accretion, and feedback in maintaining the observed mass flows.

With a measure of the kinetic energy density as a function of size-scale and position, ACES will therefore (i) test the relative importance of different energy injection mechanisms (e.g. shear) and how this in turn determines the location and magnitude of star formation \citep[e.g.,][]{KlessenGlover2016}; (ii) correlate the molecular and ionised gas properties with the known sources of feedback \citep[e.g.][]{Barnes2020} to measure the momentum and energy injection interior and exterior to clouds \citep{Kruijssen2014, Kruijssen2019, Federrath2016, Henshaw2016a}.

\subsection{Science Objective 2: Disentangling the 3D CMZ geometry}
\label{sub:sci_obj_2}

The CMZ is one of the most well-studied regions in astrophysics, with ongoing or recently completed large programs across the electromagnetic spectrum on MeerKAT, VLA, SMA, SOFIA, VLT, XMM-NEWTON, Chandra, etc., \citep{Heywood2022,Law2008, Hankins2020SOFIAGCLegacyOverview, Battersby2020, Nogueras-Lara2019, Ponti2015XMMCentralDegrees,Muno2009ChandraGCcatalog}. CMZ studies have produced many exciting results, e.g., key tests of general relativity \citep{Gravity2018}, and the centre of the Galaxy remains an optimal location for testing physics in extreme environments, e.g. searching for evidence for annihilating dark matter \citep{Murgia2020}. However, the potential of the CMZ as a laboratory of extreme physics is fundamentally limited by the fact that observations only provide a 2D projection of the complex interplay of different physical processes. The inability to determine the location and motions of gas and young stars across the CMZ is the limiting factor in interpreting the large volumes of observational data attempting to use the Galactic Centre to tackle fundamental questions such as the nature of dark matter.

A major source of contention in the literature is not knowing which clouds belong to the stream of gas that orbits the Galactic centre at a radius of $\sim$100\,pc. Current single-dish data cannot spatially resolve overlapping components of the stream structures, and are too coarse for comparison with X-ray studies which could provide line-of-sight gas locations \citep{Terrier2018}. Interferometric studies do not have the field of view or sensitivity to trace the gas in between the clouds. Our present view of clouds as `isolated islands' means we cannot distinguish whether they lie close to the centre and are in the process of feeding the next major supermassive blackhole accretion event, or lie at Galactocentric radii of $\lesssim$100\,pc and are about to form a new generation of super star clusters. 

The contiguous, large spatial dynamic range, multi-line, position–position–velocity coverage allows ACES to trace coherent gas streams across the CMZ rather than treating clouds as isolated objects. Multi‑line spectral decomposition will separate overlapping components, identify continuous kinematic structures, and compare them directly with synthetic position-position-velocity (PPV) cubes from the unified simulation network (Section 3). In regions with bright continuum and X‑ray reflection nebulae, ACES molecular absorption and recombination‑line measurements combined with existing X‑ray and radio data will place clouds in front of or behind Sgr A and the nuclear cluster, thereby breaking current degeneracies in the 3D gas geometry.

ACES' ability to trace the contiguous position-position-velocity structure, combined with the simulation framework, will build upon our evolving understanding of the 3D gas geometry \citep{Battersby2025a, Battersby2025b, Walker2025, Lipman2025} and help to break the remaining degeneracies \citep[e.g.][]{Henshaw2016a}, leading to a unified picture of the 3D gas structure with strong legacy value for future studies.

\subsection{Science Objective 3: Determine if there are preferred locations for star formation in the CMZ}
\label{sub:sci_obj_3}

Studies of extragalactic CMZs show there may be preferred locations at which gas clouds transition from quiescent to star-forming \citep{Boker2008}. In one scenario, star formation can be triggered at the contact point between the dust lanes feeding the nuclear ring and the ring itself \citep{Boker2008,Seo2013,Sormani2020}. In the CMZ, it has been argued that star formation may also be coordinated and triggered by tidal compression of clouds passing pericentre with the bottom of the Galactic gravitational potential \citep{Longmore2013b, Kruijssen2015}. The former scenario applied to the CMZ would predict star formation activity and properties of the embedded cores depend on where infalling gas collides with existing gas clouds \citep[at Sgr B2 and Sgr C located at apocentre;][]{Bally2010}. The latter predicts increasing star formation as clouds progress farther downstream from pericentre passage. 

ACES dust continuum emission will identify all dense structures expected to host individual star-forming cores across the survey footprint and measure their masses, sizes, and spatial distribution. Line diagnostics from HCO$^+$, HCN, CS, and their isotopologues will provide kinematic information and volume densities, while hot‑core and outflow tracers (e.g. vibrationally excited HC$_3$N, CH$_3$OH, COMs, and SiO) will flag sites of ongoing star formation and feedback. By placing these dense, star-forming gas structures and young stellar objects along dynamical orbits inferred from the gas kinematics, we will test whether core formation and star‑formation activity correlate with specific orbital phases (e.g. pericentre vs.~apocentre) or with the bar contact points, and hence whether CMZ star formation is globally coordinated or locally stochastic.

Thus, with a measure of the embedded star formation \citep[e.g.][]{Lu2019} crucially now complete as a function of location, and with a complete description of the gas distribution, ACES aims to distinguish between the model predictions and provide critical insight into the formation of super star clusters and the central starburst activity that plays a key role in galaxy evolution \citep[e.g.][]{Leroy2018}.

\subsection{Science Objective 4: Determine whether star formation theory predictions hold in extreme environments}
\label{sub:sci_obj_4}

Theories combining the scale-free physics of turbulence with gravity in a multi-freefall framework \citep{MacLowKlessen2004, McKeeOstriker2007, HennebelleChabrier2011} predict star formation varies strongly with the virial parameter, $\alphavir$, the turbulent Mach number, $\mathcal{M}$, and the driving mode of turbulence, $b$ \citep[e.g.,][]{FederrathKlessen2012}. Specifically, in many of the modern theories of the star formation rate (SFR) and the initial mass function (IMF) the `critical density' for star formation, $\rho_{\rm crit}$, is proportional to $\alphavir\mathcal{M}^2$ \citep{KrumholzMcKee2005,PadoanNordlund2011}, and the peak and width of the pre-stellar core mass function and the IMF are set by the sonic scale, $R_{\rm S}$, where the turbulence transitions from supersonic to subsonic speeds \citep{FederrathEtAl2021}, which may be accompanied by gravity overcoming thermal pressure \citep{Hopkins2013, Hennebelle2013}. With a measurement of the density and velocity field, ACES will provide the data to determine $\alphavir$ and $\mathcal{M}$, as well as $\rho_{\rm crit}$ and $R_{\rm S}$ with unprecedented precision and coverage, enabling stringent tests of these predictions for all potentially star-forming gas in the CMZ \citep{Bertram2015, Barnes2017}. Compared to star formation regions in the Galactic disk, against which many of the star formation theories are calibrated, the density, velocity, and temperature are significantly different in the CMZ, leading to differences in $\rho_{\rm crit}$ and $R_{\rm S}$ by about an order of magnitude, owing to differences in key environmental properties such as the radiation field and cosmic-ray ionization rate. It is therefore critical to determine the dimensionless parameters $\alphavir$, $\mathcal{M}$, and $b$, to test the theoretical predictions of star formation, and indeed some of these parameters may vary substantially in the CMZ, posing challenges for some of the theoretical models. 

The combination of dust continuum and multi‑line spectroscopy in ACES will provide estimates of column density, volume density, velocity dispersion, and hence virial parameter and turbulent Mach number for every dense structure. While these are measured from position-position-velocity data, we will apply established and calibrated techniques to estimate and recover the intrinsic three-dimensional turbulent velocity and density fluctuations \citep{BruntFederrathPrice2010a,BruntFederrathPrice2010b,StewartFederrath2022,NarayanTritsisFederrath2025}. From these, ACES can derive the critical density for collapse and the sonic scale, and compare them with the mass function of dense structures on similar scales in the CMZ and Galactic disc, together with independent estimates of the star‑formation rate from radio and infrared tracers. This enables direct tests of theoretical predictions for how the star‑formation efficiency and the characteristic dense gas mass depend on $\alphavir$, Mach number, and turbulence driving, in an environment where these parameters might differ by an order of magnitude from typical disc clouds, but are representative of high‑redshift galaxies and local starbursts.

ACES will therefore provide a new benchmark for star formation theories in extreme environments similar to those in $z \sim 1$--3 galaxies and local starbursts \citep{Kruijssen2013} at otherwise unreachable scales. With a complete census of the masses, velocities, and velocity dispersions of dense, core‑bearing structures that host high-mass star precursors, ACES will tackle fundamental open questions in star formation research in the most extreme of dynamic environments.

\section{ACES: Data Highlights and discovery potential}
\label{sec:data_highlights}

We now show a small selection of the initial ACES data and data products, with the aim of highlighting the power of ACES' combination of angular resolution, spatial dynamic range, sensitivity, spectral resolution and spectral bandwidth. These are selected to illustrate how the ACES data will be used to address the science objectives and understand how global processes set the location, intensity, and timescales for star formation and feedback in the CMZ.

\begin{figure*}
\centering
\includegraphics[trim={0 30 0 0},clip,width=1.0\textwidth]{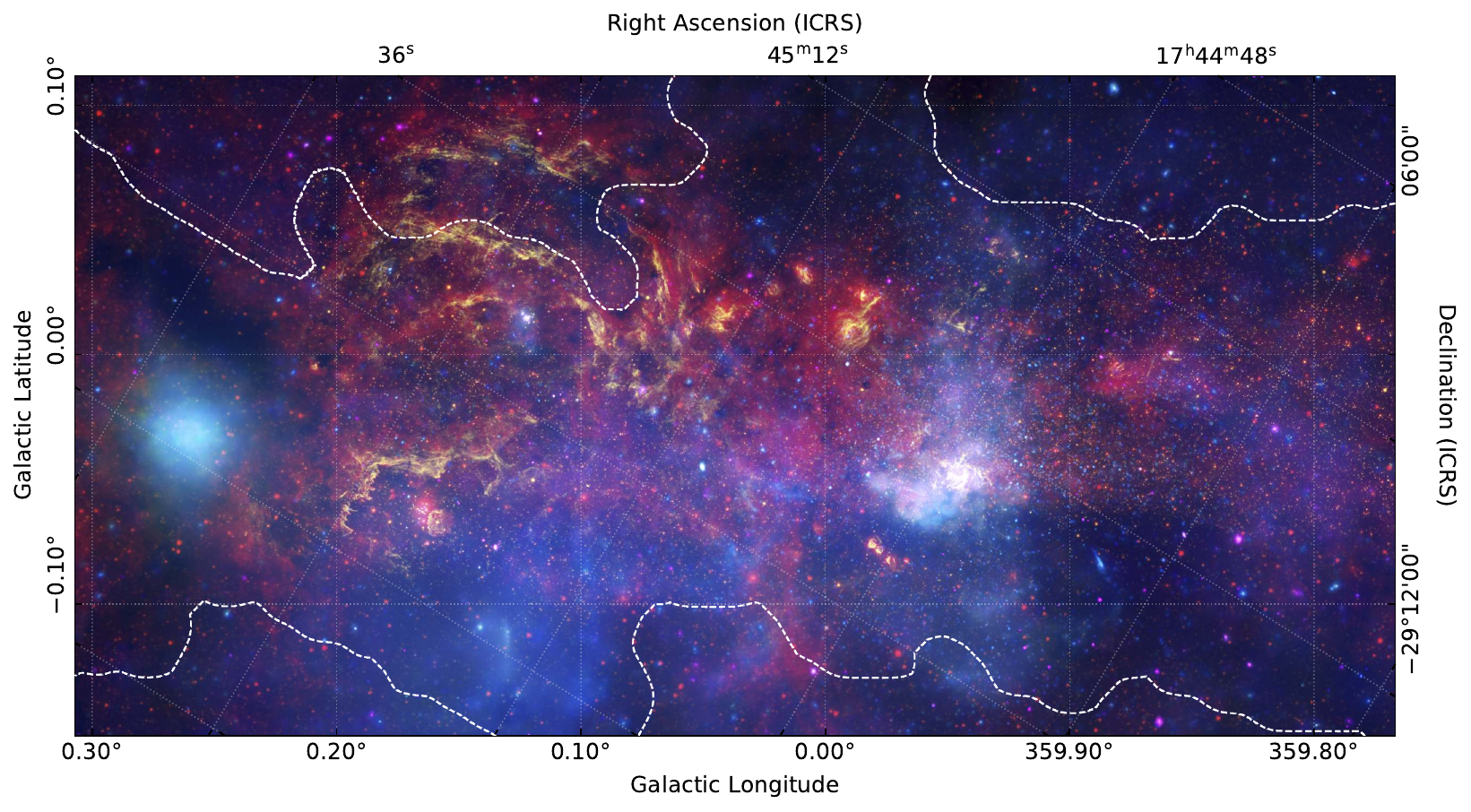} \\ \vspace{-2mm}
\includegraphics[trim={0 0 0 37},clip, width=1.0\textwidth]{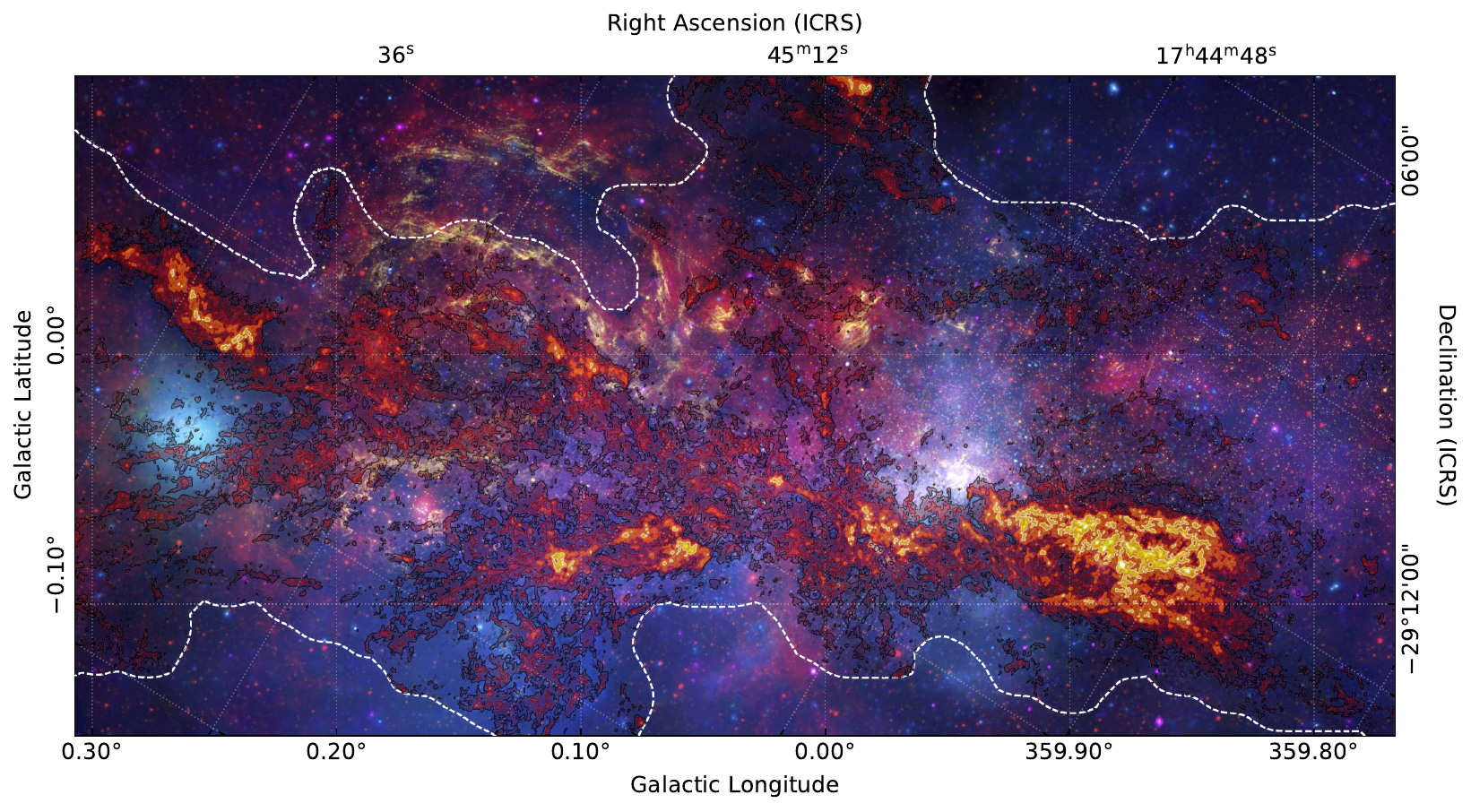} \vspace{-5mm}
\caption{Multi-wavelength view of the central region of the Galaxy. The top panel shows a composite using data from the Spitzer Space Telescope at 3.6, 4.5, 5.8, and 8.0\,$\mu$m \citep{Churchwell2009}, Hubble Space Telescope (HST) Paschen-$\alpha$ emission (F187N) and (F190N) continuum \citep{Wang2010, Dong2011}, and Chandra X-ray Observatory data in three energy bands: 1--3\,keV (soft), 3--5\,keV (medium), and 5--8\,keV (hard; see \citealp{Wang2021, Wang2022}). The bottom panel shows the same region overlaid with the ACES HNCO data from Figure \ref{fig:ACES_rgb}.
}\label{fig:3col_centre_ACES}
\end{figure*}

\begin{figure*}
\centering\includegraphics[trim={0 30 0 0},clip,width=1.0\textwidth]{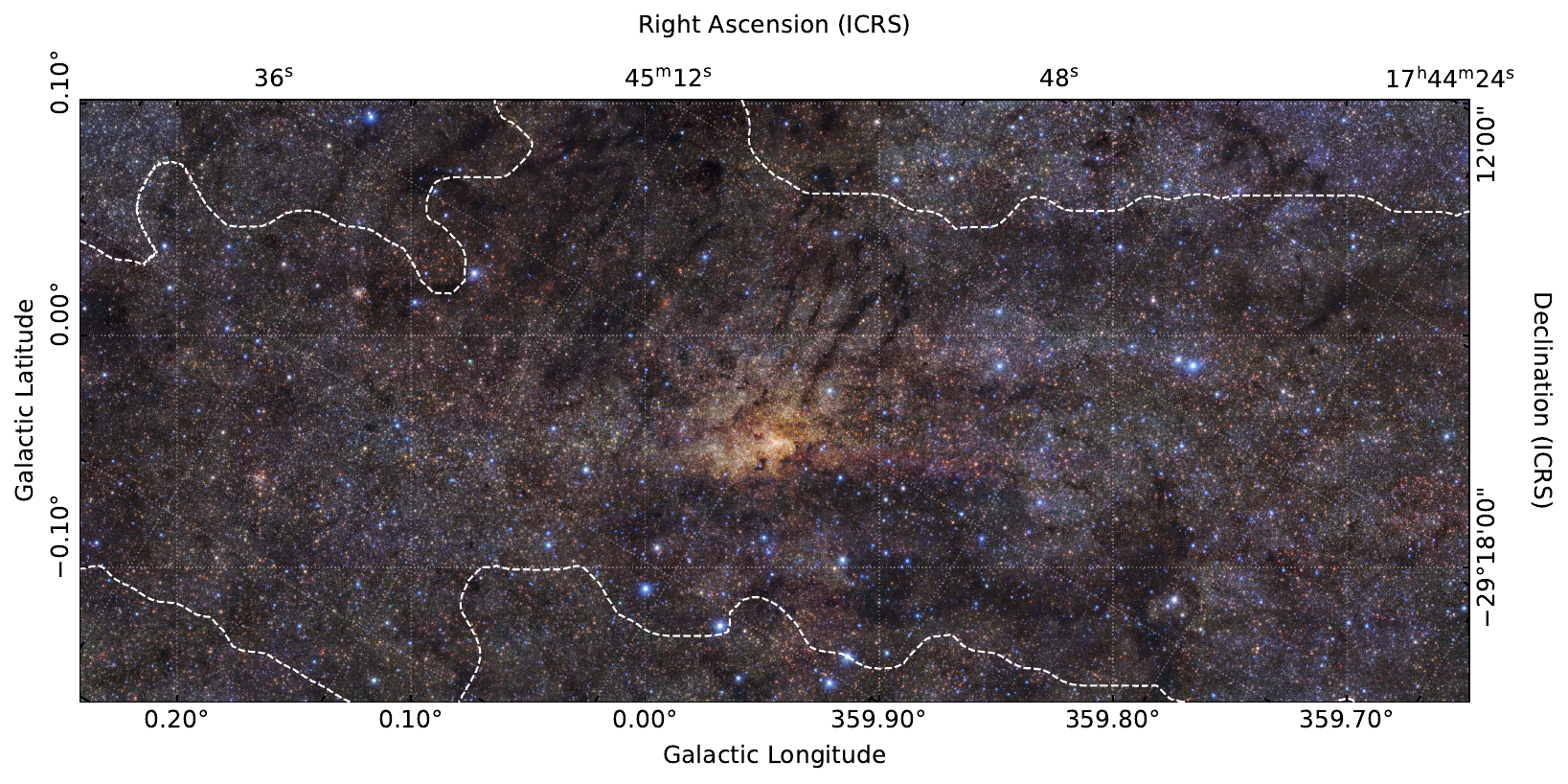} \\ \vspace{-2mm}\includegraphics[trim={0 0 0 37},clip, width=1.0\textwidth]{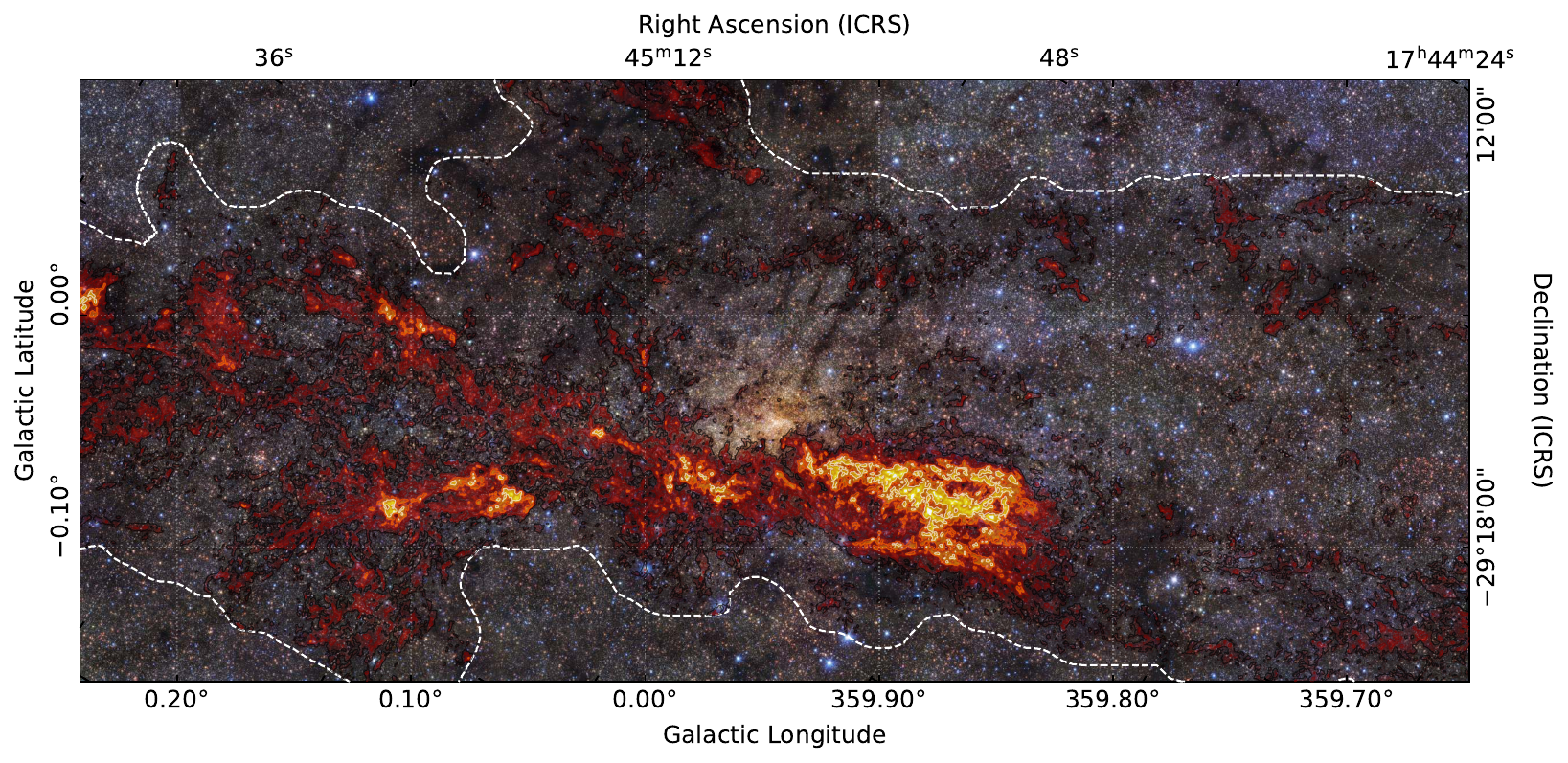} \vspace{-5mm}
\caption{Multi-wavelength view of the central region of the Galaxy. The top panel shows a 3-colour, near-IR, $JHK_s$ (VLT/HAWK-I) image as the background taken from the central $\sim15'\times36'$ of the GALACTICNUCLEUS survey \citep{Nogueras-Lara2019}. The bottom panel shows the same region overlaid with the ACES HNCO data from Figure \ref{fig:ACES_rgb}.}
\label{fig:HAWKI_centre_ACES}
\end{figure*}

\begin{figure*}
\centering\includegraphics[trim={0 28 0 0},clip,width=1.0\textwidth]{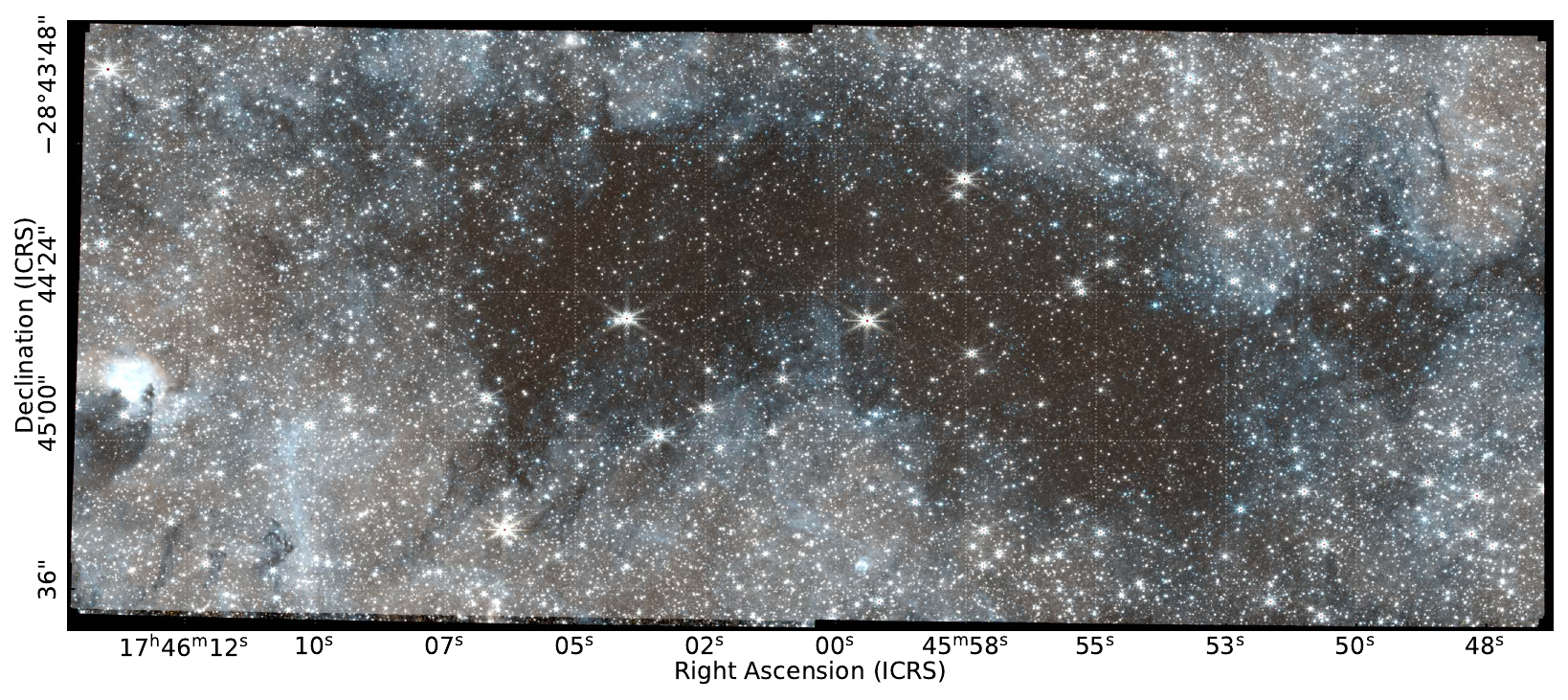} \\ \vspace{-2mm}\includegraphics[trim={0 0 0 0},clip, width=1.0\textwidth]{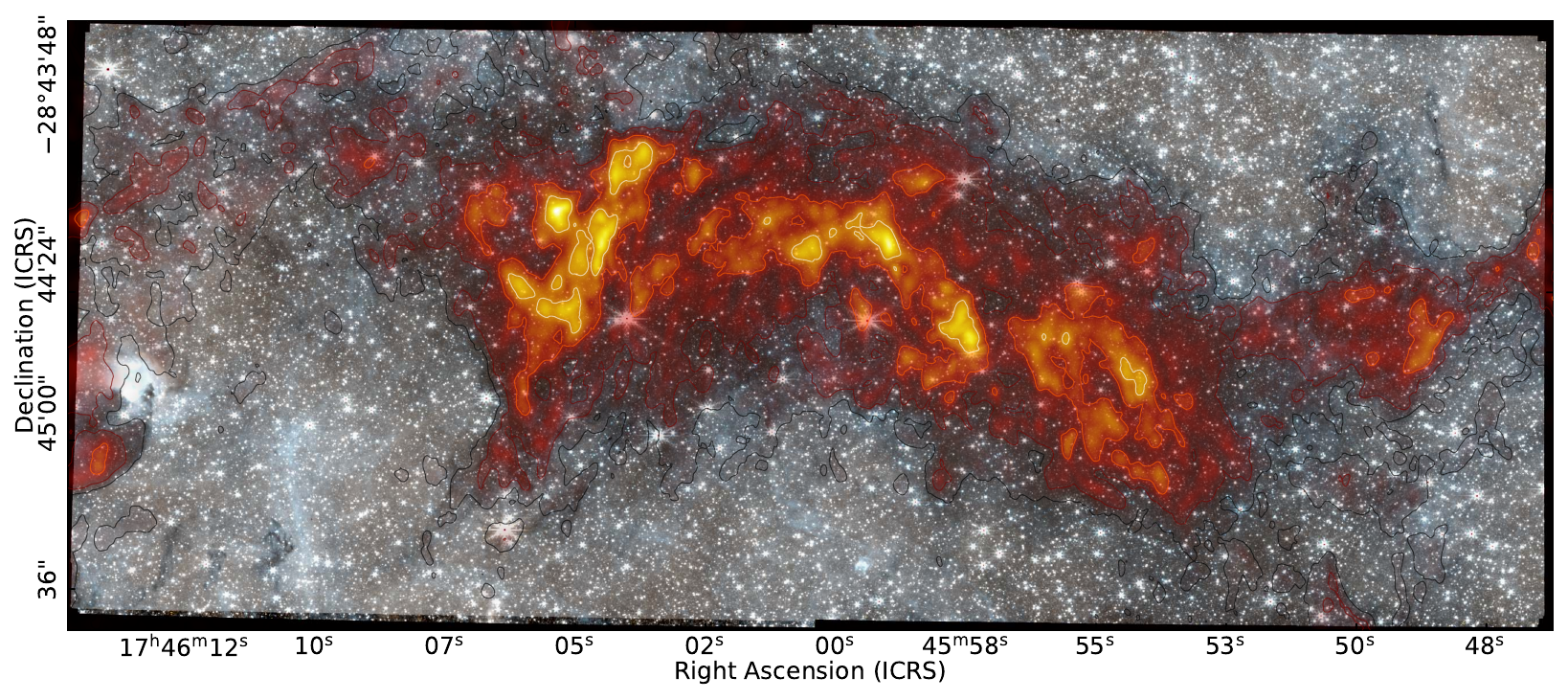} \vspace{-5mm}
\caption{Zoom in to the G0.253$+$0.016 molecular cloud, known as the `Brick'. The 3-colour image in the top and bottom panels is from James Webb Space Telescope (JWST) observations reported in \citet{Ginsburg2023}. The bottom panel shows the same region overlaid with the ACES HNCO data from Figure \ref{fig:ACES_rgb}.}
\label{fig:Brick_ACES}
\end{figure*}

Figures \ref{fig:3col_centre_ACES}, \ref{fig:HAWKI_centre_ACES}, and \ref{fig:Brick_ACES} compare the ACES HNCO data in Figure \ref{fig:ACES_rgb} with other multi-wavelength datasets. Figures \ref{fig:3col_centre_ACES} and \ref{fig:HAWKI_centre_ACES} zoom in to the very central region of the Galaxy, containing the arched filaments, the Arches and Quintuplet stellar clusters, the nuclear stellar cluster, and supermassive black hole, Sgr A$^*$. Figure \ref{fig:Brick_ACES} zooms in to the very massive (10$^5$\,M$_\odot$), compact (radius of a few pc) G0.253$+$0.016 molecular cloud, known as the `Brick'. 

These figures highlight several important early findings from the survey. Firstly, both the ACES line and continuum data contain emission at a large range of spatial scales, with a lot of complex substructure down to the resolution limit. Measuring and understanding what determines this large spatial dynamic range in emission structures and the variation as a function of velocity in different lines is a key goal of the survey. 

Secondly, there is no simple interpretation for the emission from an individual transition. For example, these figures demonstrate that HNCO does a good job of tracing the extinction regions  over much of the CMZ. As such, it is likely a good general probe of the dense gas structure. However, there are a small number of pronounced extinction features with no obvious (or only weak) HNCO emission. It is clear that a more sophisticated, multi-line interpretation is needed to understand the spatial and velocity structure of the gas than simply assigning HNCO as a ``dense gas'' or ``shock'' tracer.

\begin{figure*}
\centering\includegraphics[width=1.0\textwidth]{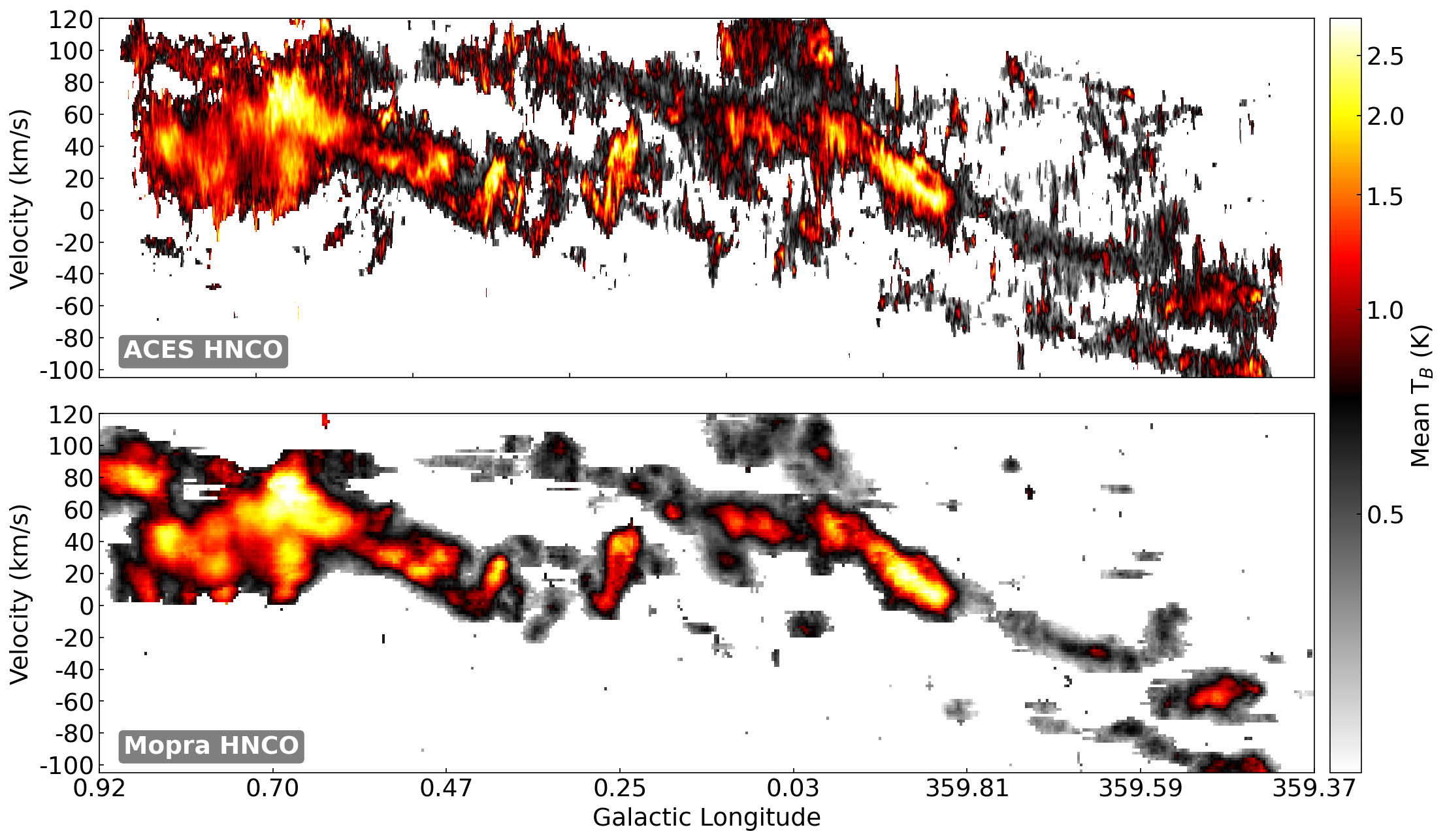} \\
\caption{Masked HNCO emission from ACES [top] and the Mopra CMZ survey \citep{Jones2012} [bottom] summed along Galactic Latitude to show emission velocity as a function of Galactic Longitude. See \citet{Walker2025_ACESIII} for more details on the masking techniques used to create these maps.}
\label{fig:HNCO_PV}
\end{figure*}

\subsection{Chemistry and Physical Modeling}
\label{sub:chem_phys_modelling}

\subsubsection{Molecular benchmarking with known sources}
\begin{figure*}
\centering\includegraphics[width=1.0\textwidth]{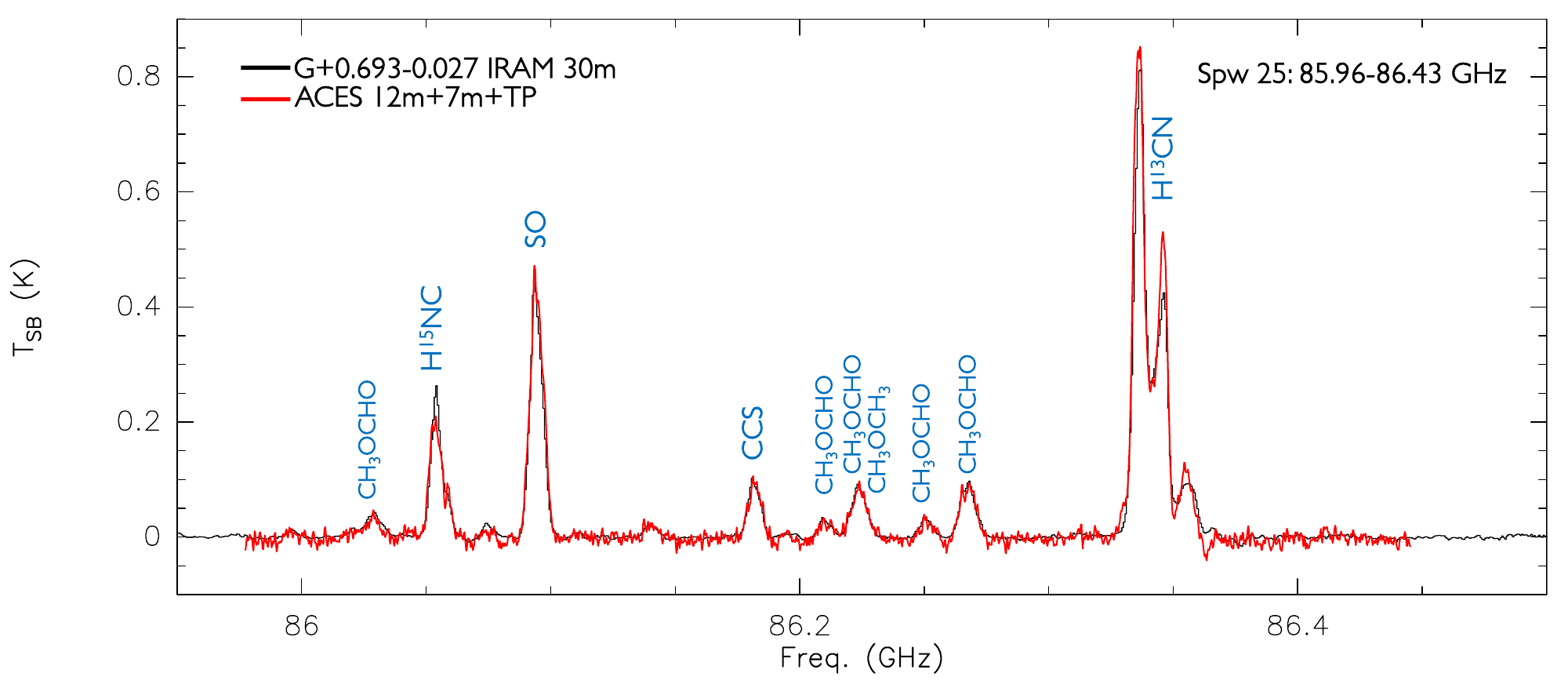} \\ \vspace{-2mm}\includegraphics[clip, width=1.0\textwidth]{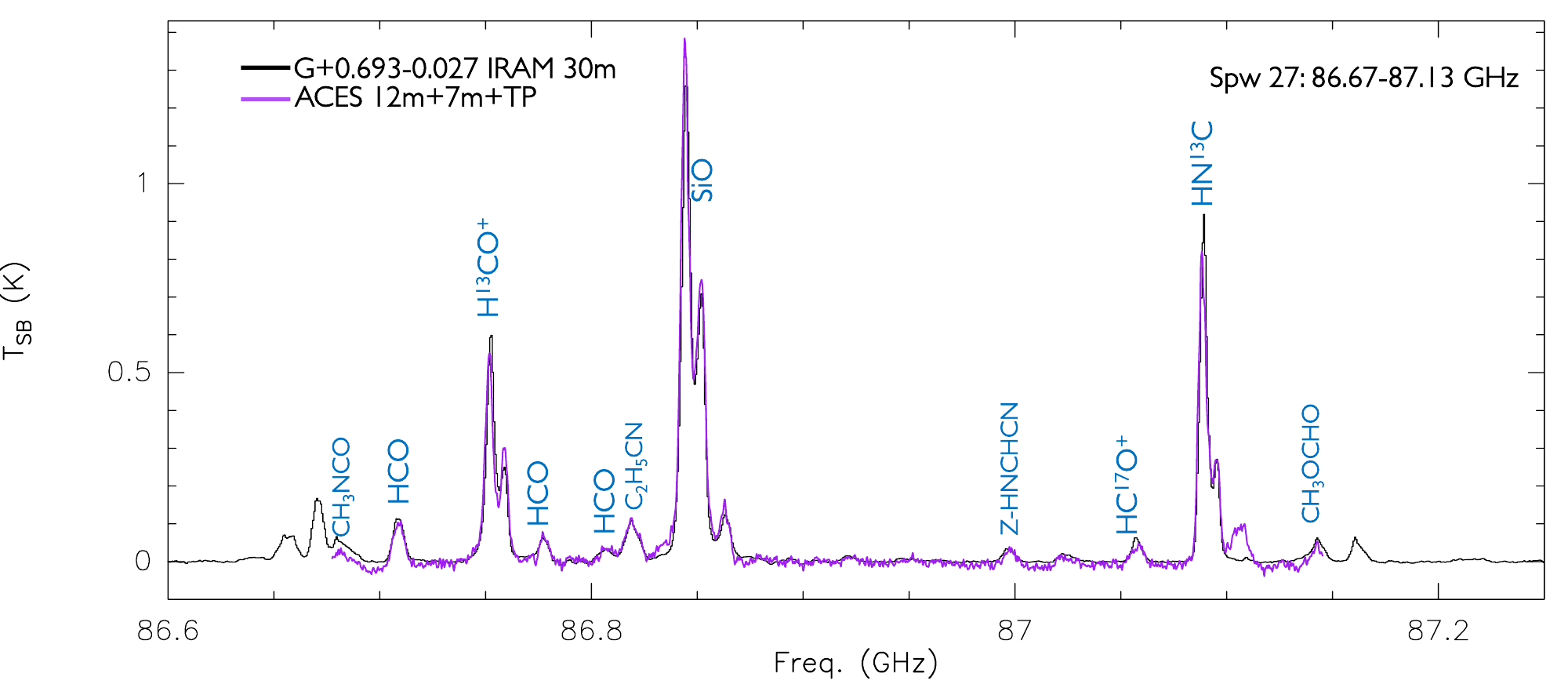}
\caption{Comparison between the ACES 12m+7m+TP (in red) SPW 25
 (\textit{top panel}) and SPW 27 (\textit{bottom panel}) with single-dish spectra obtained with the IRAM 30m (in black) towards the G$+$0.693 molecular cloud. The blue labels indicate the name of the molecules whose transitions are detected in the observed spectral window. Paper III discusses the technical details and resulting robustness of the single-dish data combination in detail.}\label{fig:line_id1}
\end{figure*}

\begin{figure*}
\centering\includegraphics[width=1.0\textwidth]{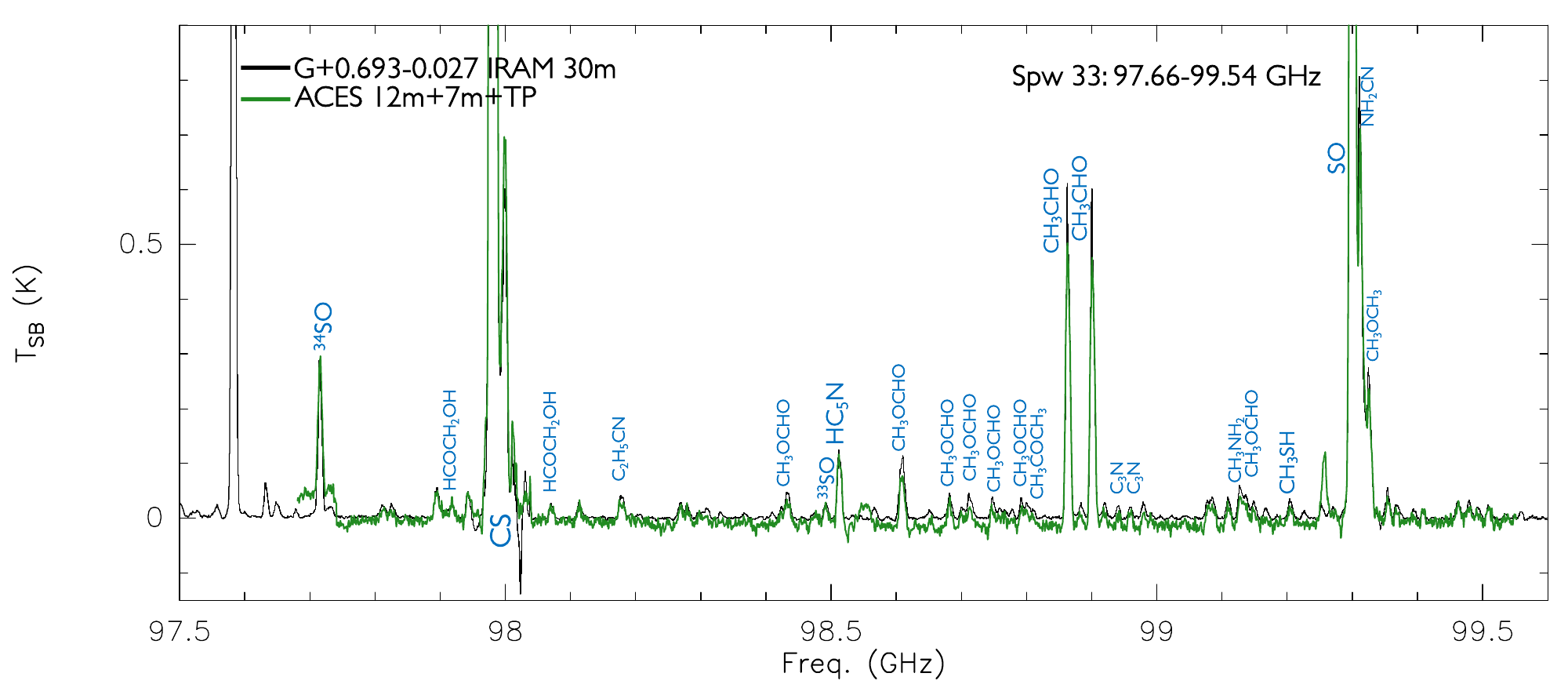} \\ \vspace{-2mm}\includegraphics[clip, width=1.035\textwidth]{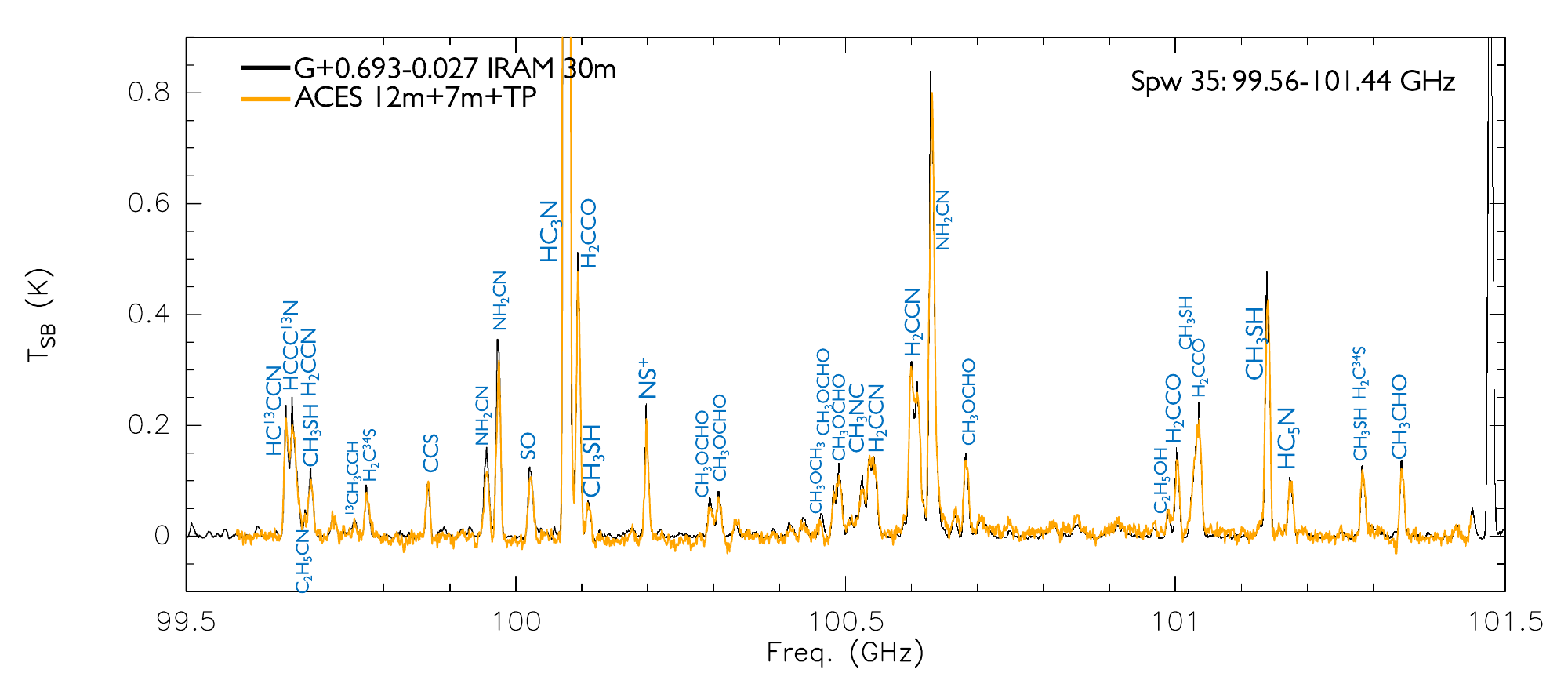}
\caption{Same as Figure \ref{fig:line_id1} for ACES spws 33 (\textit{top panel}) and 35 (\textit{bottom panel}).}\label{fig:line_id2}
\end{figure*}

ACES covers 6 spectral windows in ALMA band 3 at 3 mm \citep[see Table 1 in][ for full details]{Ginsburg2025_ACESII}. Two of these are high spectral resolution spectral windows ($\sim$0.2 km s$^{-1}$) devoted to kinematic studies using HCO$^{+}$(1--0) and HNCO(4--3) rotational transitions, and will not be discussed in the following. The other four spectral windows cover observed frequencies 85.96--86.43, 86.67--87.13, 97.66--99.54, and 99.56--101.44\,GHz at 1.7 to 2.9\,\kms\ spectral resolution. These observed frequencies cover rotational transitions of many molecular species (see Figures \ref{fig:line_id1} and \ref{fig:line_id2}), allowing for the determination of the physical and chemical properties of the molecular and ionized gas structures from the largest scales ($\sim$100 pc) to the dense $\sim$0.05\,pc structures that host star‑forming cores.

The brightest molecular lines in the ACES Band 3 setup are high‑dipole‑moment species such as HCO$^+$, HCN, and CS that are traditionally used as “dense‑gas tracers” because their ground‑state transitions have high formal critical densities compared to low‑J CO \citep[e.g.][]{shirley2015}. However, wide‑field surveys of nearby molecular clouds have demonstrated that the $J=1-0$ emission from HCN and HCO$^+$ is readily excited in extended, moderate‑density gas and that a substantial fraction of their luminosity arises from translucent cloud envelopes rather than from the highest‑density, gravitationally bound cores \citep[][]{pety2017, kauffmann2017, tafalla2021, tafalla2023}. We use the term “dense‑gas tracers” for this family of high‑dipole lines as a shorthand, while explicitly treating their emission as a probe of comparatively high column‑density gas rather than as a one‑to‑one tracer of gas above a fixed volume‑density threshold.

In addition to these high‑dipole species, the ACES spectral windows include molecules that are commonly associated with shocks (e.g. SiO, HNCO, SO), classic photon-dominated region (PDR) tracers such as HCO, radio recombination lines (H40$\alpha$, H50$\beta$) tracing ionised gas, and hot‑core tracers such as the v$_7=1$ and v$_6 = 1$ vibrationally excited lines of HC$_3$N. As already seen for complex organic molecules (COMs) in the CMZ, these associations need to be interpreted with care: COM emission can arise either from massive hot cores \citep[e.g., Sgr B2 (N);][with high excitation temperatures, T$_{\rm ex}$$\geq$150\,K]{belloche2013} or from intermediate‑density gas processed by large‑scale shocks \citep[with much lower excitation temperatures, T$_{\rm ex}$$\leq$15\,K;][]{zeng2018,zeng2020}. Similar caveats apply to other “standard” tracer labels in the extreme CMZ environment (see below).

The physical conditions in the CMZ differ substantially from those in the nearby disc clouds where these tracer identifications were originally calibrated. On $10-100$\,pc scales, the gas is on average denser [n(H$_2$) $\gtrsim 10^4$\,cm$^{-3}$], warmer (T$_{\rm K} \gtrsim 50-100$\,K), and more turbulent (typical line widths $\gtrsim 10-20$\,km\,s$^{-1}$) than in the Galactic disc, and is permeated by strong magnetic fields and an enhanced cosmic‑ray ionisation rate \citep[e.g.][]{Morris1996, ao2013, Ginsburg2016, Krieger2017, Henshaw2023}. In such an environment, radiative trapping, elevated temperatures, and large velocity gradients reduce the effective densities required to excite high‑dipole molecules \citep[][]{shirley2015}, while high‑temperature chemistry, increased ionisation, and widespread low‑velocity shocks can strongly modify molecular abundances and line ratios \citep[e.g.][]{harada2010, zeng2018, zeng2020}. As a consequence, species that in disc clouds often act as relatively selective tracers of dense cores, shocks, or PDRs are expected to be more widespread in the CMZ, and their emission reflects a convolution of density, temperature, and environmental effects. A core goal of ACES is therefore to calibrate the behaviour of these “standard” tracers under CMZ conditions by combining multi‑line spectroscopy, dust‑based column densities, and dedicated astrochemical modelling tailored to the CMZ \citep[Section 5.1.2;][]{Dutkowska2025}.

Molecular gas in the CMZ is highly turbulent, with supersonic linewidth and shocked material (e.g. \citealt{elmegreen2004,MacLowKlessen2004}). A typical example of a chemically rich shocked region in the Galactic Center, without on-going star formation or protostellar activity, is the G$+$0.693$-$0.027 (G$+$0.693 hereafter) molecular cloud, located in the Sgr B2 region, 55\arcsec\ northeast from Sgr B2(N). In the last 5 years more than 20 new interstellar molecules have been discovered towards this molecular cloud using very sensitive single-dish radio telescope observations \citep[e.g.,][]{rivilla2021,jimenez-serra2022,sanandres2024,sanz-novo2024,sanz-novo2025}. G$+$0.693 has been recently discovered to harbor a prestellar condensation, hypothesized to be formed from the shock produced by a cloud-cloud collision, which could be the cradle of the next generation of stars in the CMZ \citep{zeng2020,colzi2022,colzi2024}. ACES high angular resolution observations will allow us to properly map the molecular emission towards this condensation.

In Figures \ref{fig:line_id1} and \ref{fig:line_id2} we show the observed spectra towards G$+$0.693 using single-dish observations obtained with the IRAM 30m radiotelescope (see \citealt{colzi2024} for more details on the observations) at the frequencies of SPWs 25, 27, 33, 35. We superimposed the spectra extracted from the ACES cube (12m + ACA 7m + total power) towards the coordinates of the source $\alpha_{\rm J2000}= 17^{\rm h}47^{\rm m}22^{\rm s}$ and $\delta_{\rm J2000} = -28^\circ 21^{\prime}27\arcsec$ and from a region matching the size of the IRAM 30m beam at the observed frequency (29\arcsec\;at 86 GHz).

The match between the single-dish and interferometric observations is very good, indicating the reliability of the data reduction and merging of the different ALMA configurations used to recover the more extended flux. There are some small differences ($<20\%$ in synthesized beam temperature, $T_{\rm SB}$), mainly due to the continuum subtraction, which is not straightforward in the CMZ due to the multiple gas components, especially in the Sgr B2 region (see Paper II by Ginsburg et al., in prep.). In the IRAM 30m spectra, some absorption features appear in very strong lines, like CS at 98 GHz, which are not present in the ACES data. These are due to the subtraction of the off-position while taking the single-dish data.

This comparison with single-dish data shows the potential of ACES data in detecting many molecular species, from simple molecules such as SO, SiO and CS
to more complex ones such as CH$_{3}$CHO, CH$_{3}$SH, and NH$_{2}$CN.

\begin{table*}
\caption{COMs and COM precursors covered within the ACES frequency setup\label{tab:coms}}
\begin{center}
\begin{tabular}{cccc}
\hline\hline
\multicolumn{4}{c}{COMs families}  \\  \hline
    O-bearing & N-bearing & S-bearing & C-bearing \\
\hline
CH$_3$OH, CH$_{3}$CHO, &   C$_{2}$H$_{5}$CN, CH$_{3}$NH$_{2}$,  & CH$_{3}$SH & $^{13}$CH$_3$CCH\\
 CH$_{3}$OCHO, H$_{2}$CCO, & H$_{2}$CCN, CH$_{3}$NC,   & & \\
\textbf{C}H$_{3}$OCH$_{3}$, CH$_{3}$COCH$_{3}$,  & NH$_{2}$CN, HC$_5$N,& & \\
C$_{2}$H$_{5}$OH, HCOCH$_{2}$OH &CH$_3$NCO, CH$_3$C$_3$N  & \\
\hline\hline
\end{tabular}
\end{center}
\end{table*}

Typical molecular excitation temperatures ($T_{\rm ex}$) towards G$+$0.693 are of the order of 3-20 K (e.g. \citealt{zeng2018}), and thus only the low energy transitions are excited, making the observed spectra less crowded than towards a typical hot molecular core \citep[like the Sgr B2 hot cores,][]{belloche2013,sanchez2017,moeller2025} with $T_{\rm ex}>$ 100 K. Taking into account these high excitation conditions, in hot molecular cores other molecular species, apart from those already shown in G$+$0.693, can be detected, such as H$_{2}$CO, CH$_{3}$OH, CH$_{3}$C$_{3}$N, CH$_{3}$NCO, HC$_{3}$N vibrationally excited, HC(O)NH$_{2}$. Furthermore, radio recombination lines H40$\alpha$ and H50$\beta$ are also included in the ACES setup and are well known tracers of ionized gas, e.g., towards HII regions.

In Table~\ref{table-simplemolecules} we list the frequencies of the most common rotational transitions of simple molecules present in the ACES setup. In addition,  in Table \ref{tab:coms}, we also list the COMs and COM precursors that we have covered within the frequency setup of ACES. Note we do not list specific transitions for COMs since their excitation and presence in the spectra is highly dependent on the excitation condition of the cloud targeted.

\subsubsection{Astrochemical models}
\begin{figure*}
\centering\includegraphics[width=1\textwidth]{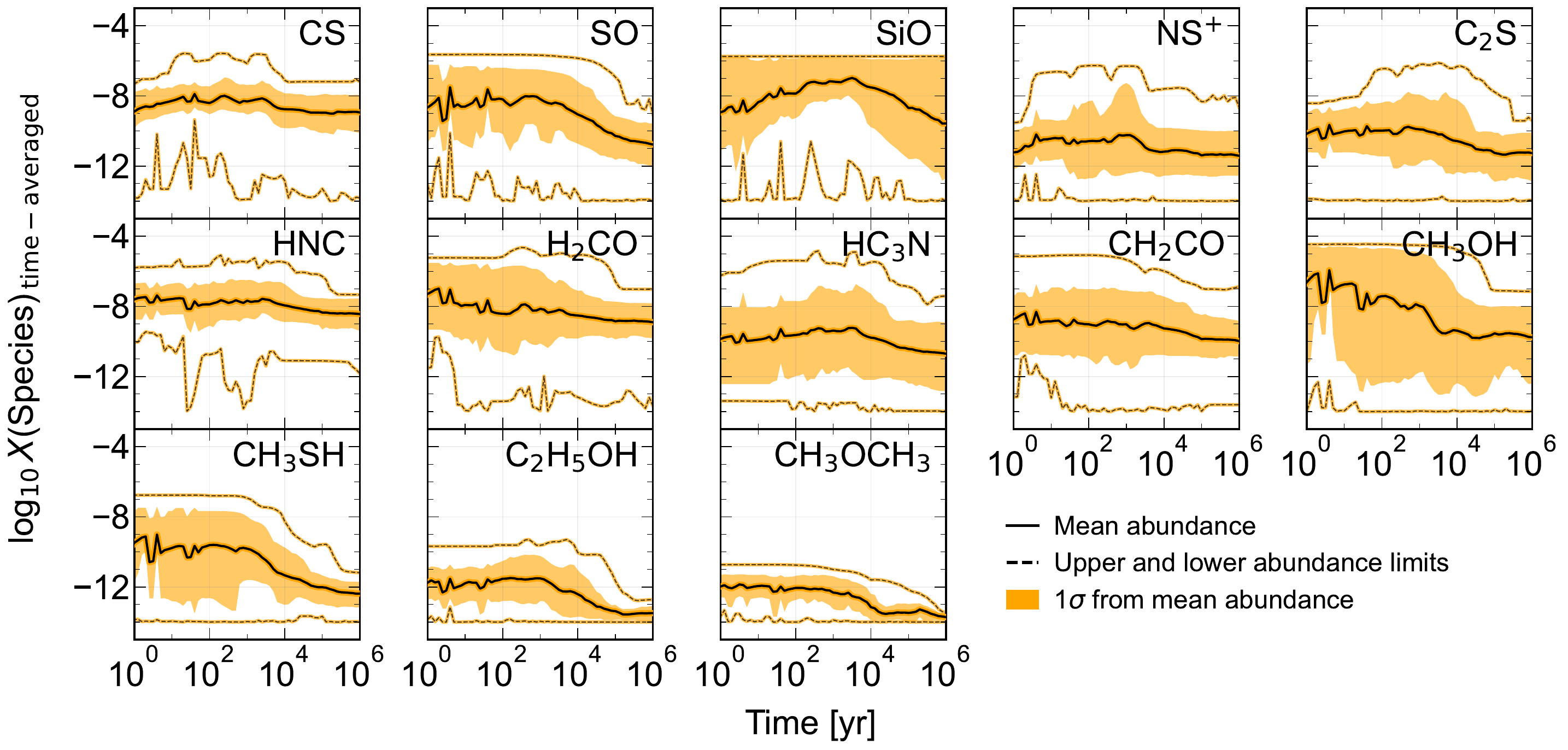}
\caption{Time-averaged chemical evolution of some species for the C-shock models. Only gas-phase abundances above the observable threshold ($X >$ 10$^{-14}$) are considered.  CH$_{3}$OH is the most dominant species during the shock phase, while SiO takes over in the post-shock phase. Adapted from \citet{Dutkowska2025}.  }\label{fig:chemical_model}
\end{figure*}

Astrochemical modelling is vital to interpret the ACES spectral line observations. As part of a wider coordinated effort from multiple research groups, we have recently completed the astrochemical modeling of sources with typical CMZ conditions: protostellar objects designed to resemble cores like those in the Sgr B2 cloud, and low-to-medium velocity C-type shocks like those occurring during cloud-cloud collisions (e.g., as observed towards G$+$0.693). These models account for key CMZ characteristics including enhanced cosmic-ray ionization rates and elevated dust temperatures. Full details of these grids of models computed using UCLCHEM (\citealt{holdship2017}) are given in \citet{Dutkowska2025}.

In Figure \ref{fig:chemical_model} we show the average chemical abundances derived for the C-shock models as a function of time for a subset of modelled species. During the shock phase ($\lesssim$ 10$^{4}$ years), the rise in temperature and density begins almost immediately, leading to general gas phase abundance enhancements. However, in the post-shock stage (after 10$^{4}$ years) the gas phase abundances generally decrease as the gas and dust cool back to their initial temperatures and the density increases, promoting the freeze-out of species onto icy mantles of dust grains.

The $\sim10^4$ year timescale reflects the short-lived nature of the shocks in the considered parameter space. For densities of $10^4-10^6\;\text{cm}^{-3}$ and shock velocities in the range of $10-40\;\text{km\,s}^{-1}$, virtually all shocks occur on timescales of $\lesssim$ 10$^{4}$ years, with higher densities leading to shorter timescales. Considering that Figure \ref{fig:chemical_model} averages all shock models, the apparent scatter and early decline of abundances of certain species around 10$^{4}$ years reflects this range of shock events rather than a single shock scenario \citep[for a detailed description of shock evolution, see Figure 1 in][]{Dutkowska2025}.

Predictions for dedicated chemical models over a range of physical conditions expected in the CMZ environment can be found in \citet{Dutkowska2025}. These and subsequent models will be vital to interpret the ACES data in the wide range of different physical environments across the Galactic Centre.

\subsection{ACES Early Science Demonstration: measuring feedback energy and momentum transfer}

A key aspect of ACES Science Objective 1 is to understand how energy and momentum from stellar feedback couples to the surrounding gas and drives the mass flows as a function of size scale and location. Initial science results from ACES data have revealed a striking case of feedback-driven cloud disruption in the CMZ. \citet{Nonhebel2024} present a detailed analysis of the M0.8$-$0.2 ring, a prominent shell-like molecular cloud complex in the southeastern extension of Sgr\,B2. Leveraging ACES high angular resolution, sensitivity, spatial dynamic range, and spectral coverage, \citet{Nonhebel2024} determine the complex is an expanding ring of dense, shocked gas with a radius of 6.1\,pc, expansion velocity of $\sim$21\,\kms, and a mass approaching $10^6$\,M$_\odot$. The derived kinetic energy and momentum ($>$10$^{51}$ erg, $>$10$^7$\,M$_\odot$\,\kms) point to an energetic origin, with detailed radiative transfer modeling and multi-wavelength comparison ruling out early-stage stellar feedback or multiple clustered supernovae (SNe). Instead, \citet{Nonhebel2024} propose a single high-energy hypernova explosion, likely originating from a runaway massive star, as the most plausible driver of the structure. The inferred energetics challenge canonical models of SNR evolution \citep[e.g.][]{Thornton1998, Kim2015a} and suggest the need for modified feedback prescriptions in high-density environments like the CMZ.

\subsection{ACES Early Science Demonstration: Disentangling the 3D CMZ geometry}

\citet{Sofue2025} highlight the power of ACES data in separating different velocity structures in the kinematically complex Galactic Centre environment to address Science Objective 2 and understand the 3D structure of the CMZ. Using ACES molecular line data alongside Nobeyama and ASTE observations, \citet{Sofue2025} identify and characterize a nested system of six spiral arms (Arms I–VI) within the Central Molecular Zone (CMZ). These `arms' appear as coherent, tilted longitude-velocity (LV) ridges in CS and HCN lines, and are interpreted as a sequence of nearly circular, inclined ring-like structures centered on Sgr~A*. A new, faint inner arm (Arm V) and the circumnuclear disk (CND; Arm VI) are highlighted, with the ionised minispiral around Sgr~A* proposed to constitute an innermost seventh arm.

\subsection{ACES Discovery Potential: Expect the Unexpected}

Any survey, like ACES, which provides orders of magnitude improvements in observational capabilities has the potential to make unexpected discoveries. As part of ACES data quality assessment and early science, \citet{Ginsburg2025} discovered a compact millimeter-bright source near the 50~km\,s$^{-1}$ cloud, designated G0.02467$-$0.0727. This object, coined the \textit{Millimeter Ultra-Broad Line Object} (MUBLO), exhibits exceptionally broad molecular line emission (full width at half maximum (FWHM)~$\approx$~160\,km\,s$^{-1}$) from SO, SO$_2$, and CS transitions, yet remains remarkably compact ($\lesssim10^4$\,au in diameter). The MUBLO line profiles are well-fit by a Gaussian with no indication of self-absorption or wings, and there is a mild but significant spatial offset between red- and blue-shifted emission, hinting at internal velocity structure.

The continuum emission has a steep spectral index ($\alpha \approx 3.25$), suggesting optically thin dust with a mass of $\sim$50\,M$_\odot$ assuming standard CMZ dust properties. Despite the large inferred mass and velocity dispersion, the object shows no sign of high-velocity shocks (e.g., no SiO emission), and no counterpart is detected at infrared, centimeter, or X-ray wavelengths. Its gas appears cold ($T_\mathrm{LTE} \approx 13$\,K), chemically unusual (SO/SiO~$\gg$~100), and isolated in projection.

Multiple physical scenarios are evaluated---including a protostellar or explosive outflow, evolved star, stellar merger remnant, high-velocity compact cloud, or gas bound to an intermediate-mass black hole---but none fully explain the observed properties. The MUBLO appears to be an unprecedented and potentially new class of molecular object in the CMZ, highlighting ACES’ ability to uncover rare and enigmatic sources through high-resolution, wide-area mapping.

\section{Conclusions}
\label{sec:conclusions}

The ACES Large Program delivers a contiguous ALMA Band 3 view of  all gas dense enough to form stars in the inner $\sim$100\,pc of the Milky Way, at $\sim$0.05\,pc resolution and with uniform, multi-line coverage. By design, this dataset links the global bar-driven inflow and $\sim$100\,pc stream to the dense structures expected to host individual protostellar cores. This paper describes the survey strategy, tracer selection, and unified simulations framework, and presents first results that already illustrate how ACES can be used to quantify the CMZ baryon cycle.

The first data products show that it is possible to recover both compact and extended emission across the full CMZ with a single, homogeneous setup. The combination of 12m, 7m, and total-power data reliably reproduce single-dish spectra over a chemically rich line forest toward G$+$0.693, while delivering $\sim1.5"$ resolution across the entire footprint. This validates the imaging strategy and confirms that ACES can be used both for core-scale work and for global mass and momentum budgets on $\sim100$\,pc scales.

The initial ACES results probe how commonly used millimetre tracers behave under CMZ conditions. HNCO emission broadly follows the dust extinction morphology over the CMZ but shows departures in some regions, demonstrating that it is a good structural tracer of dense gas but not a simple one-to-one proxy for “shocks” or a fixed density threshold. Likewise, high-dipole molecules traditionally labelled as “dense-gas tracers” are  excited over a range of column densities. The dedicated astrochemical grids developed for ACES indicate that abundance patterns and line ratios in the CMZ reflect a convolution of density, temperature, shocks, and irradiation, rather than any single controlling parameter.

Early science results already quantify how feedback reshapes the CMZ gas. The expanding M0.8$-$0.2 ring contains nearly 10$^6$\,M$_\odot$ of shocked material, with kinetic energy and momentum exceeding 10$^{51}$\,erg and 10$^7$\,M$_\odot$\,km\,s$^{-1}$, respectively, pointing to a single very energetic explosion, plausibly a hypernova, as its origin. This demonstrates that rare, high-energy events can disrupt massive clouds and inject substantial momentum into the CMZ, and that feedback prescriptions calibrated on isolated supernovae may underpredict the impact of such events in dense galactic nuclei.

The ACES kinematic maps are already sharpening our view of the three-dimensional gas structure. By separating overlapping components in position–position–velocity space, early results show how the data provide a coherent geometric framework for interpreting ongoing and future multi-wavelength studies of the Galactic Centre.

Finally, the survey has already uncovered objects that fall well outside standard categories. The Millimeter Ultra-Broad Line Object (MUBLO) near the 50\,km\,s$^{-1}$ cloud combines very broad (FWHM $\sim$160\,km\,s$^{-1}$) molecular lines, cold excitation temperatures, and a compact, massive dust core, without associated infrared, centimetre, or X-ray emission. Its properties are not matched by known classes of star-forming regions, evolved stars, or compact remnants. The existence of such an object within the first ACES fields demonstrates that the combination of wide-area coverage and high sensitivity in the CMZ is likely to reveal additional rare or previously unrecognised phases of molecular gas and compact sources.

In summary, even at this early stage, ACES has (i) demonstrated that a single, homogeneous ALMA setup can deliver a complete, flux-recovered view of dense gas across the inner 100\,pc; (ii) shown that standard molecular tracers in the CMZ require careful interpretation in terms of the physical conditions they probe; (iii) provided direct measurements of feedback-driven energy and momentum injection into $\sim$10$^6$\,M$_\odot$ clouds; (iv) refined the emerging picture of the CMZ as a system of nested, inclined gas streams; and (v) revealed at least one compact source that defies any existing classification. As the full analysis progresses, combining these data with the unified simulations framework will allow us to measure how mass, momentum, and energy flow through the CMZ, and to test star-formation and feedback theories in an environment that closely resembles the central regions of high-redshift galaxies and local starbursts.

\section*{Data Availability}
All data products and associated documentation can be found at \url{https://almascience.org/alma-data/lp/aces}.

All code and data processing issues are available at the public GitHub repository here: \url{https://github.com/ACES-CMZ/reduction_ACES}.

\section*{Acknowledgements}

The paper was instigated and led by ACES Principal Investigator Steven Longmore, together with the management team John Bally, Ashley Barnes, Cara Battersby, Laura Colzi, Adam Ginsburg, Jonathan Henshaw, Paul Ho, Izaskun Jiménez-Serra, J. M. Diederik Kruijssen, Elisabeth Mills, Maya Petkova, Mattia Sormani, Robin Tress, Daniel Walker, and Jennifer Wallace. The data reduction working group is coordinated by Adam Ginsburg, Daniel Walker, and Ashley Barnes, and includes Nazar Budaiev, Laura Colzi, Savannah Gramze, Pei-Ying Hsieh, Desmond Jeff, Xing Lu, Jaime Pineda, Marc Pound, and Álvaro Sánchez-Monge, together with more than 30 additional team members who contributed to the data reduction effort. All team members contributed to reading and commenting on the manuscript.

This paper makes use of the following ALMA data: ADS/JAO.ALMA\#2021.1.00172.L. ALMA is a partnership of ESO (representing its member states), NSF (USA) and NINS (Japan), together with NRC (Canada), NSTC and ASIAA (Taiwan), and KASI (Republic of Korea), in cooperation with the Republic of Chile. The Joint ALMA Observatory is operated by ESO, AUI/NRAO and NAOJ.
The authors are grateful to the staff throughout the ALMA organisation, particularly those at the European ALMA Regional Centre, the Joint ALMA Observatory, and the UK ALMA Regional Centre Node, for their extensive support, which was essential to the success of this challenging Large Program.

COOL Research DAO \citep{cool_whitepaper} is a Decentralized Autonomous Organization supporting research in astrophysics aimed at uncovering our cosmic origins.

C.\ Battersby  gratefully  acknowledges  funding  from  National  Science  Foundation  under  Award  Nos. 2108938, 2206510, and CAREER 2145689, as well as from the National Aeronautics and Space Administration through the Astrophysics Data Analysis Program under Award ``3-D MC: Mapping Circumnuclear Molecular Clouds from X-ray to Radio,” Grant No. 80NSSC22K1125.

L.C., I.J-S., and V.M.R. acknowledge support from grant no. PID2022-136814NB-I00 by the Spanish Ministry of Science, Innovation and Universities/State Agency of Research MICIU/AEI/10.13039/501100011033 and by ERDF, UE. I.J.-S. also acknowledges funding from the ERC Consolidator grant OPENS (project number 101125858) funded by the European Union.
V.M.R. also acknowledges the grant RYC2020-029387-I funded by MICIU/AEI/10.13039/501100011033 and by "ESF, Investing in your future", and from the Consejo Superior de Investigaciones Cient{\'i}ficas (CSIC) and the Centro de Astrobiolog{\'i}a (CAB) through the project 20225AT015 (Proyectos intramurales especiales del CSIC); and from the grant CNS2023-144464 funded by MICIU/AEI/10.13039/501100011033 and by “European Union NextGenerationEU/PRTR.

M.C.\ gratefully acknowledges funding from the DFG through an Emmy Noether Research Group (grant number CH2137/1-1).

RSK and SCOG acknowledge financial support from the European Research Council via ERC Synergy Grant ``ECOGAL'' (project ID 855130),  from the German Excellence Strategy via the Heidelberg Cluster ``STRUCTURES'' (EXC 2181 - 390900948), and from the German Ministry for Economic Affairs and Climate Action in project ``MAINN'' (funding ID 50OO2206).  RSK also thanks the 2024/25 Class of Radcliffe Fellows for highly interesting and stimulating discussions.

MCS acknowledges financial support from the European Research Council under the ERC Starting Grant ``GalFlow'' (grant 101116226) and from Fondazione Cariplo under the grant ERC attrattivit\`{a} n. 2023-3014

CF acknowledges funding provided by the Australian Research Council (Discovery Project grants~DP230102280 and~DP250101526), and the Australia-Germany Joint Research Cooperation Scheme (UA-DAAD).

MRK acknowledges funding from the Australian Research Council through Laureate Fellowship FL220100020.

A.S.-M.\ acknowledges support from the RyC2021-032892-I grant funded by MCIN/AEI/10.13039/501100011033 and by the European Union `Next GenerationEU’/PRTR, as well as the program Unidad de Excelencia María de Maeztu CEX2020-001058-M, and support from the PID2023-146675NB-I00 (MCI-AEI-FEDER, UE).

The research of KMD and SV is funded by the European Research Council (ERC) Advanced Grant MOPPEX 833460.vii.

The authors acknowledge UFIT Research Computing for providing computational resources and support that have contributed to the research results reported in this publication.

AG acknowledges support from the NSF under grants CAREER 2142300, AAG 2008101, and particularly AAG 2206511 that supports the ACES large program.

D.L.W gratefully acknowledges support from the UK ALMA Regional Centre (ARC) Node, which is supported by the Science and Technology Facilities Council [grant numbers ST/Y004108/1 and ST/T001488/1].

Q. Z. gratefully acknowledges the support from the National Science Foundation under Award No. AST-2206512, and the Smithsonian Institute FY2024 Scholarly Studies Program.

FHL acknowledges support from the ESO Studentship Programme, the Scatcherd European Scholarship of the University of Oxford, and the European Research Council’s starting grant ERC StG-101077573 (`ISM-METALS').

J.Wallace gratefully acknowledges funding from National Science Foundation under Award Nos. 2108938 and 2206510.

J.K. is supported by the Royal Society under grant number RF\textbackslash ERE\textbackslash231132, as part of project URF\textbackslash R1\textbackslash211322.

FNL gratefully acknowledges financial support from grant PID2024-162148NA-I00, funded by MCIN/AEI/10.13039/501100011033 and the European Regional Development Fund (ERDF) “A way of making Europe”, from the Ramón y Cajal programme (RYC2023-044924-I) funded by MCIN/AEI/10.13039/501100011033 and FSE+, and from the Severo Ochoa grant CEX2021-001131-S, funded by MCIN/AEI/10.13039/501100011033.

\bibliographystyle{mnras}
\bibliography{ACES_overview}

\section*{Author Affiliations}
\printaffiliation{ljmu}{Astrophysics Research Institute, Liverpool John Moores University, IC2, Liverpool Science Park, 146 Brownlow Hill, Liverpool L3 5RF, UK}
\printaffiliation{COOL}{Cosmic Origins Of Life (COOL) Research DAO, \href{https://coolresearch.io}{https://coolresearch.io}}
\printaffiliation{colorado}{Center for Astrophysics and Space Astronomy, Department of Astrophysical and Planetary Sciences, University of Colorado, Boulder, CO 80389, USA}
\printaffiliation{eso}{European Southern Observatory (ESO), Karl-Schwarzschild-Stra{\ss}e 2, 85748 Garching, Germany}
\printaffiliation{uconn}{Department of Physics, University of Connecticut, 196A Auditorium Road, Unit 3046, Storrs, CT 06269, USA}
\printaffiliation{cab_csic}{Centro de Astrobiolog\'ia (CAB), CSIC-INTA, Ctra. de Ajalvir Km. 4, 28850, Torrej\'on de Ardoz, Madrid, Spain}
\printaffiliation{uflorida}{Department of Astronomy, University of Florida, P.O. Box 112055, Gainesville, FL 32611, USA}
\printaffiliation{mpia}{Max Planck Institute for Astronomy, K\"onigstuhl 17, D-69117 Heidelberg, Germany}
\printaffiliation{iaa_taipei}{Academia Sinica Institute of Astronomy and Astrophysics, Astronomy-Mathematics Building, AS/NTU No.1, Sec. 4, Roosevelt Rd, Taipei 10617, Taiwan}
\printaffiliation{kansas}{Department of Physics and Astronomy, University of Kansas, 1251 Wescoe Hall Drive, Lawrence, KS 66045, USA}
\printaffiliation{chalmers}{Space, Earth and Environment Department, Chalmers University of Technology, SE-412 96 Gothenburg, Sweden}
\printaffiliation{clap}{Como Lake centre for AstroPhysics (CLAP), DiSAT, Universit\`{a} dell’Insubria, via Valleggio 11, 22100 Como, Italy}
\printaffiliation{iop_epfl}{Institute of Physics, Laboratory for Galaxy Evolution and Spectral Modelling, EPFL, Observatoire de Sauverny, Chemin Pegasi 51, 1290 Versoix, Switzerland}
\printaffiliation{ukarcnode}{UK ALMA Regional Centre Node, Jodrell Bank Centre for Astrophysics, The University of Manchester, Manchester M13 9PL, UK}
\printaffiliation{mpifr_bonn}{Max-Planck-Institut f\"ur Radioastronomie, Auf dem H\"ugel 69, 53121 Bonn, Germany}
\printaffiliation{kau_jeddah}{Astronomy Department, Faculty of Science, King Abdulaziz University, P.O. Box 80203, Jeddah 21589, Saudi Arabia}
\printaffiliation{princeton}{Princeton}
\printaffiliation{armagh}{Armagh}
\printaffiliation{nrao}{National Radio Astronomy Observatory, 520 Edgemont Road, Charlottesville, VA 22903, USA}
\printaffiliation{mpe}{Max-Planck-Institut f\"ur extraterrestrische Physik, Giessenbachstrasse 1, D-85748 Garching, Germany}
\printaffiliation{ita_heidelberg}{Universit\"at Heidelberg, Zentrum f\"ur Astronomie, Institut f\"ur Theoretische Astrophysik, Albert-Ueberle-Str. 2, 69120 Heidelberg, Germany}
\printaffiliation{unknown_affil}{???}
\printaffiliation{uva}{Dept. of Astronomy, University of Virginia, Charlottesville, Virginia 22904, USA}
\printaffiliation{upenn}{Department of Physics and Astronomy, 209 S. 33rd Street, University of Pennsylvania, PA 19104, USA}
\printaffiliation{leiden}{Leiden Observatory, Leiden University, P.O. Box 9513, 2300 RA Leiden, The Netherlands}
\printaffiliation{anu}{Research School of Astronomy and Astrophysics, Australian National University, 233 Mt Stromlo Road, Stromlo ACT 2611, Australia}
\printaffiliation{iaa_csic}{Instituto de Astrof\'{i}sica de Andaluc\'{i}a, CSIC, Glorieta de la Astronom\'ia s/n, 18008 Granada, Spain}
\printaffiliation{malta}{Institute of Space Sciences \& Astronomy, University of Malta, Msida MSD 2080, Malta}
\printaffiliation{jbca}{Jodrell Bank Centre for Astrophysics, The University of Manchester, Manchester M13 9PL, UK}
\printaffiliation{cassaca}{Chinese Academy of Sciences South America Center for Astronomy, National Astronomical Observatories, CAS, Beijing 100101, China}
\printaffiliation{ucn}{Instituto de Astronom\'ia, Universidad Cat\'olica del Norte, Av. Angamos 0610, Antofagasta, Chile}
\printaffiliation{csic_gen}{CSIC}
\printaffiliation{heidelberg_univ}{University of Heidelberg}
\printaffiliation{northwestern}{North Western}
\printaffiliation{shao}{Shanghai Astronomical Observatory, Chinese Academy of Sciences, 80 Nandan Road, Shanghai 200030, P. R. China}
\printaffiliation{naoj}{National Astronomical Observatory of Japan, 2-21-1 Osawa, Mitaka, Tokyo 181-8588, Japan}
\printaffiliation{ias}{Institute for Advanced Study, 1 Einstein Drive, Princeton, NJ 08540, USA}
\printaffiliation{ucl}{Department of Physics and Astronomy, University College London, Gower Street, London, UK}
\printaffiliation{mit}{Haystack Observatory, Massachusetts Institute of Technology, 99 Millstone Road, Westford, MA 01886, USA}
\printaffiliation{izw_heidelberg}{Universit\"at Heidelberg, Interdisziplin\"ares Zentrum f\"ur Wissenschaftliches Rechnen, Im Neuenheimer Feld 205, 69120 Heidelberg, Germany}
\printaffiliation{cfa}{Center for Astrophysics | Harvard \& Smithsonian, 60 Garden Street, Cambridge, MA 02138, USA}
\printaffiliation{radcliffe}{Elizabeth S. and Richard M. Cashin Fellow at the Radcliffe Institute for Advanced Studies at Harvard University, 10 Garden Street, Cambridge, MA 02138, USA}
\printaffiliation{uw_madison}{Department of Astronomy, University of Wisconsin-Madison, Madison, WI, 53706, USA}
\printaffiliation{uw_seattle}{Department of Astronomy, Box 351580, University of Washington, Seattle, WA 98195, USA}
\printaffiliation{ari_heidelberg}{Astronomisches Rechen-Institut, Zentrum f\"{u}r Astronomie der Universit\"{a}t Heidelberg, M\"{o}nchhofstra\ss e 12-14, D-69120 Heidelberg, Germany}
\printaffiliation{naoc_key}{State Key Laboratory of Radio Astronomy and Technology, A20 Datun Road, Chaoyang District, Beijing, 100101, P. R. China}
\printaffiliation{ucas_astro}{School of Astronomy and Space Sciences, University of Chinese Academy of Sciences, No. 19A Yuquan Road, Beijing 100049, People’s Republic of China}
\printaffiliation{eso_chile}{European Southern Observatory, Alonso de C\'ordova, 3107, Vitacura, Santiago 763-0355, Chile}
\printaffiliation{jao}{Joint ALMA Observatory, Alonso de C\'ordova, 3107, Vitacura, Santiago 763-0355, Chile}
\printaffiliation{ipm}{Institute for Research in Fundamental Sciences (IPM), School of Astronomy, Tehran, Iran}
\printaffiliation{ucla}{Department of Physics \& Astronomy, University of California, Los Angeles, Los Angeles, CA 90095-1547, USA}
\printaffiliation{standrews_eso}{St Andrews/ESO}
\printaffiliation{keio_univ}{Department of Physics, Faculy of Science and Technology, Keio University, 3-14-1 Hiyoshi, Kohoku-ku, Yokohama, Kanagawa 223-8522, Japan}
\printaffiliation{inaf_gen}{INAF}
\printaffiliation{nanjing}{School of Astronomy and Space Science, Nanjing University, 163 Xianlin Avenue, Nanjing 210023, P.R.China}
\printaffiliation{nanjing_key}{Key Laboratory of Modern Astronomy and Astrophysics (Nanjing University), Ministry of Education, Nanjing 210023, P.R.China}
\printaffiliation{umd}{University of Maryland, Department of Astronomy, College Park, MD 20742-2421, USA}
\printaffiliation{ulaserena}{Departamento de Astronom\'ia, Universidad de La Serena, Ra\'ul Bitr\'an 1305, La Serena, Chile}
\printaffiliation{ice_csic}{Institute of Space Sciences (ICE, CSIC), Campus UAB, Carrer de Can Magrans s/n, 08193, Bellaterra (Barcelona), Spain}
\printaffiliation{ieec}{Institute of Space Studies of Catalonia (IEEC), 08860, Barcelona, Spain}
\printaffiliation{gbo}{Green Bank Observatory, P.O. Box 2, Green Bank, WV 24944, USA}
\printaffiliation{nice}{Nice}
\printaffiliation{utokyo}{Institute of Astronomy, The University of Tokyo, Mitaka, Tokyo, 181-0015, Japan}
\printaffiliation{bologna}{ESO/Bologna}
\printaffiliation{insubria}{University of Insubria}
\printaffiliation{tra_bonn}{Transdisciplinary Research Area (TRA) ‘Matter’/Argelander-Institut f\"ur Astronomie, University of Bonn, Bonn, Germany}
\printaffiliation{umass}{Department of Astronomy, University of Massachusetts, Amherst, MA 01003, USA}
\printaffiliation{kiaa_pku}{Kavli Institute for Astronomy and Astrophysics, Peking University, Beijing 100871, People's Republic of China}

\appendix

\section{ACES -- an open, transparent, community-wide approach to producing data products and papers}
\label{sec:data_products}

The ACES collaboration is open to anyone interested in contributing to the project, with membership approved through the management board. The science objectives are facilitated by tasks within four different work packages, led by members of the management team, dedicated to (i) data reduction, (ii) producing the ACES fundamental measurements, (iii) chemistry and physical modelling, and (iv) the unified simulation suite. To ensure complete transparency and fair attribution of credit for key technical tasks, progress within the work packages is tracked using the ACES CMZ public GitHub project (\url{https://github.com/ACES-CMZ}). For example, all of the data reduction steps from the delivery of each individual scheduling block, through to production of the final mosaics, can be tracked through separate issues on the data reduction repository (\url{https://github.com/ACES-CMZ/reduction\_ACES}). This provides a permanent record of the wide range of technical challenges that needed to be overcome, and the dedicated effort from a large number of ACES team members, ALMA staff, contact scientists, technical staff, etc., required to solve the data reduction challenges and produce the final, full-field mosaics. Given the unique challenges associated with the different data products, a full description of the specific data reduction steps are given in the corresponding Papers II to V.

\section{Data Release Summary}
\label{sec:data_release_summary}


\subsection{ACES data products and file naming}
\label{subsec:data_products}

All ACES data products are released through the ALMA Science Archive (ASA) and
follow a uniform file-naming scheme that is consistent across the continuum and
spectral-line papers (Papers~II--V). The hierarchy of products mirrors the
standard ALMA project structure, with Member ObsUnitSet (MOUS) products for
individual fields and arrays, and Group ObsUnitSet (GOUS) products for combined
regions and the full CMZ mosaics.

\subsubsection{Member-level products (per field, per array)}

For each of the 45 ACES regions we release member-level products corresponding
to the images or cubes for the individual ALMA 12-m and 7-m arrays prior to any
array combination or mosaicking.

The generic filename template for member-level products is
\texttt{member.uid\_\_\_A001\_\{mous-id\}.\allowbreak lp\_slongmore.\allowbreak\{field\_coords\}.\allowbreak\{array\}.\allowbreak\{freq\_descriptor\}.\allowbreak\{product\}.fits}

where:
\begin{itemize}
  \item \verb|{mous-id}| is the 12-m or 7-m MOUS ID for the field.
  \item \verb|{field_coords}| encodes the central field coordinates (e.g.
        \verb|G000.073+0.184|).
  \item \verb|{array}| specifies the array (\verb|12m| or \verb|7m|; for
        continuum we distribute 12-m only).
  \item \verb|{freq_descriptor}| encodes the spectral setup:
    \begin{itemize}
      \item for continuum products (Paper~II), this is a centre frequency and
            bandwidth (e.g. \verb|99.6GHz_bw3.8GHz|),
      \item for spectral-line cubes (Papers~III, IV and V), this is the frequency
            range associated with a given SPW (e.g. \verb|86.0-86.4GHz|,
            \verb|97.7-99.5GHz|), as listed in Walker \& ACES Team (2025).
    \end{itemize}
  \item \verb|{product}| specifies the type of data product.
\end{itemize}

For continuum member products (Paper~II), \verb|{product}| has the form
\verb|cont.{contsuffix}| where \verb|{contsuffix}| combines the multi-term
deconvolution Taylor term (\verb|tt0|, \verb|tt1|) with the image type.
Distributed continuum products include:
\begin{itemize}
  \item \verb|image|, \verb|image.pbcor| (primary-beam corrected),
  \item \verb|model|,
  \item \verb|residual|.
\end{itemize}

For spectral-line member products (Papers~III, IV and V), \verb|{product}| is
typically
\verb|cube.pbcor|
for the primary-beam corrected image cube per SPW and per array. These cubes
reflect the ACES reprocessing and therefore differ from the original ALMA-delivered
products; the stand-alone total power (TP) cubes are not re-released, as we use
the TP data as available in the ASA.

A full description of the member-level continuum products is provided in
Paper~II, while the member-level spectral-line cubes and their frequency
coverage are described in Papers~III, IV and V (see also Appendix~A of
Walker \& ACES Team 2025).

\subsubsection{Group-level products: array-combined cubes per region}

For each region and SPW we also provide group-level array-combined cubes that
merge the interferometric and single-dish data. These products combine multiple
MOUSs and arrays and thus sit at the GOUS level.

The generic template for the array-combined regional cubes is
\texttt{group.uid\_\_\_A001\_X1590\_X30a9.\allowbreak lp\_slongmore.\allowbreak\{field\_coords\}.\allowbreak\{arrays\}.\allowbreak\{freq\_descriptor\}.\allowbreak cube.pbcor.fits}

where:
\begin{itemize}
  \item \verb|{field_coords}| and \verb|{freq_descriptor}| follow the same
        conventions as for the member-level products,
  \item \verb|{arrays}| specifies the array combination used, most commonly
        \verb|12m7mTP| for combined 12-m + 7-m + TP data.
\end{itemize}

These products are described in detail in Papers~III, IV and V, which summarise the
beam properties, pixel scales, and noise characteristics per region and SPW.

\subsubsection{Group-level products: full CMZ mosaics}

ACES also delivers a set of full, contiguous mosaics covering the entire survey
footprint. These are GOUS-level products that combine all relevant MOUSs into a
single CMZ-wide data set.

There are two broad classes of CMZ mosaics:

\paragraph{Continuum mosaics (Paper II).}

These are built for multiple continuum frequency setups and are provided for
12-m-only data and, where applicable, combined 12-m + GBT/MUSTANG products.

\paragraph{Spectral-line mosaics (Papers III--V).}

These are constructed only for selected bright and widespread lines rather than
for every SPW, in order to avoid very large cubes dominated by blank channels.
They include HNCO and HCO$^{+}$ (Paper~III), their isotopologues and related
dense-gas/shock tracers (Papers~IV and V), and other key diagnostics such as
CS~(2--1), hydrogen recombination lines, and complex organic tracers. The exact
list of lines and their SPW assignments is given in the line papers (see
Tables~1 in Papers~III--V).

The filename template for all CMZ-wide mosaics and their associated high-level
products is
\texttt{group.uid\_\_\_A001\_X1590\_X30a9.\allowbreak lp\_slongmore.\allowbreak cmz\_mosaic.\allowbreak\{arrays\}.\allowbreak\{tracer\}.\allowbreak\{suffix\}}

with:
\begin{itemize}
  \item \verb|{arrays}| describing the array combination used, e.g.
    \begin{itemize}
      \item \verb|12m| or \verb|12mGBT-MUSTANG| for continuum-only products
            (Paper~II),
      \item \verb|12m7mTP| for fully combined line mosaics (Papers~III--V),
      \item \verb|12m7mMopra| for combinations including external single-dish
            data (Paper~III),
      \item \verb|12m7m| for interferometric-only line mosaics (Paper~III).
    \end{itemize}
  \item \verb|{tracer}| specifying either
    \begin{itemize}
      \item \verb|cont| for continuum mosaics (Paper~II), or
      \item a molecule / transition name written to match the filenames, e.g.
            \verb|HNCO|, \verb|HCOplus|, \verb|HCOplus_noTP|, \verb|H13CN|,
            \verb|H13COplus|, \verb|HC15N|, \verb|HN13C|, \verb|HC3N|,
            \verb|CS21|, \verb|H40a|, \verb|CH3CHO|, \verb|SiO21|,
            \verb|SO21|, \verb|SO32| (Papers~III--V).
    \end{itemize}
  \item \verb|{suffix}| defining the particular product type (see below).
\end{itemize}

\subsubsection{Derived maps and diagnostic products}

In addition to the primary cubes and images, we distribute a uniform suite of
derived and diagnostic products for both continuum and selected spectral lines.

For the continuum mosaics (Paper~II), the key high-level products are:
\begin{itemize}
  \item Primary-beam corrected continuum mosaics:
    \begin{itemize}
      \item \verb|pbcor.fits|
    \end{itemize}
  \item Corresponding uncertainty maps:
    \begin{itemize}
      \item \verb|maskedrms.pbcor.fits|
    \end{itemize}
  \item Spectral-index diagnostics for the aggregate continuum bandwidth:
    \begin{itemize}
      \item \verb|alpha.fits|, \verb|alpha.maskedrms.fits| (automated spectral
            index and its uncertainty),
      \item \verb|alpha_manual.fits|, \verb|alpha_manual.maskedrms.fits|
            (manually curated masks and corresponding uncertainties).
    \end{itemize}
\end{itemize}

For the spectral-line mosaics (Papers~III--V), we provide:
\begin{itemize}
  \item Full-resolution line cubes:
    \begin{itemize}
      \item \verb|cube.pbcor.fits|
    \end{itemize}
  \item Spatially (and, where appropriate, spectrally) downsampled cubes to
        facilitate analysis of large-scale structure:
    \begin{itemize}
      \item \verb|cube.downsampled_spatially.pbcor.fits| (spatial smoothing to
            a common beam and spatial rebinning),
      \item \verb|cube.downsampled_spatially_and_spectrally.pbcor.fits|
            (where supplied; see Paper~III for details).
    \end{itemize}
  \item Two-dimensional maps derived from the full-resolution cubes:
    \begin{itemize}
      \item \verb|integrated_intensity.fits| -- masked integrated intensity
            maps,
      \item \verb|mad_std.fits| -- noise maps,
      \item \verb|peak_intensity.fits| -- peak intensity maps,
      \item \verb|velocity_at_peak_intensity.fits| -- velocity at the intensity
            peak for each pixel.
    \end{itemize}
  \item Position--velocity diagnostics along Galactic longitude and latitude:
    \begin{itemize}
      \item \verb|PV_l_max.fits|, \verb|PV_l_mean.fits| -- $\ell$--$v$
            diagrams using maximum or mean intensity along Galactic latitude,
      \item \verb|PV_b_max.fits|, \verb|PV_b_mean.fits| -- $b$--$v$ diagrams
            using maximum or mean intensity along Galactic longitude.
    \end{itemize}
\end{itemize}

The construction, masking, and error estimation procedures for these derived
products are described in detail in the line papers: HNCO and HCO$^{+}$ in
Paper~III, and the broader set of lines in the intermediate-band spectral
windows in Papers~IV and V. The continuum imaging strategy and spectral-index
mapping are discussed in Paper~II.





\bsp
\label{lastpage}
\end{document}